\documentclass[reprint,NumberedRefs]{Acta_SI_mine4arXiv}

\usepackage{amsmath}
\usepackage{algpseudocode}
\usepackage{graphicx}
\usepackage[utf8]{inputenc}
\usepackage{enumitem}

\usepackage{hyphenat}
\begin{document}

\title[Comparison of auditory models]{A comparative study of eight human auditory models of monaural processing}
\author{Alejandro Osses Vecchi}
\affiliation{Laboratoire des systèmes perceptifs, Département d'études cognitives,
École Normale Supérieure, PSL University, CNRS,
Paris, France}

\author{Léo Varnet}
\affiliation{Laboratoire des systèmes perceptifs, Département d'études cognitives,
École Normale Supérieure, PSL University, CNRS,
Paris, France}

\author{Laurel H. Carney}
\affiliation{Departments of Biomedical Engineering and Neuroscience, University of Rochester, Rochester, NY, USA}

\author{Torsten Dau}
\affiliation{Hearing Systems Section, Department of Health Technology, Technical University of Denmark, Lyngby, Denmark}

\author{Ian C. Bruce}
\affiliation{Department of Electrical and Computer Engineering, McMaster University, Hamilton, ON, Canada}

\author{Sarah Verhulst}
\affiliation{Hearing Technology group, WAVES, Department of Information Technology, Ghent University, Ghent, Belgium}

\author{Piotr~Majdak}
\affiliation{Acoustics Research Institute, Austrian Academy of Sciences, Vienna, Austria}

\preprint{Osses et al., Acta}

\date{\today}

\begin{abstract}
A number of auditory models have been developed using diverging approaches, either physiological or perceptual, but they share comparable stages of signal processing, as they are inspired by the same constitutive parts of the auditory system. We compare eight monaural models that are openly accessible in the Auditory Modelling Toolbox. We discuss the considerations required to make the model outputs comparable to each other, as well as the results for the following model processing stages or their equivalents: Outer and middle ear, cochlear filter bank, inner hair cell, auditory nerve synapse, cochlear nucleus, and inferior colliculus. The discussion includes a list of recommendations for future applications of auditory models. 
\end{abstract}


\maketitle

\begin{figure*}
    \centering
    \includegraphics[width=0.85\textwidth]{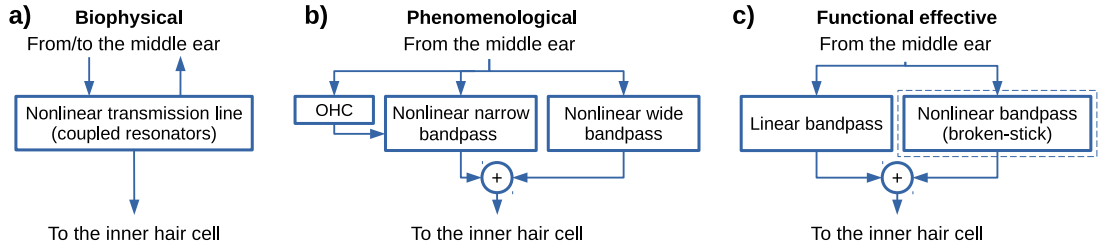}\\
    \vspace{-12pt}
    \caption{Model families used in this study. The families are defined by the level of detail in simulating the cochlear processing and are sorted by their complexity from left to right. a)\,Biophysical models using a nonlinear transmission line that contains resonating stages coupled by the cochlear fluid. b)\,Phenomenological models using nonlinear filters dynamically controlled by an outer-hair-cell (OHC) model. c)\,Functional effective models using linear filters, optionally combined with static nonlinear filters.}
    \label{fig:model-families}
\end{figure*}

\section{Introduction}
\label{sec:01-intro}

Computational auditory models reflect our fundamental knowledge about hearing processes and have been accumulated during decades of research~\cite[e.g.,][]{meddis_computational_2010}. Models are used to derive conclusions, reproduce findings, and develop future applications. Usually, models are built in stages that reflect basic parts of the auditory system, such as cochlear filtering, mechanoneural interface, and neural processing, by applying signal-processing methods such as bandpass filtering or envelope processing~\cite{Dau2008}. Models of monaural processing often form a basis for binaural models  \cite[e.g.,][]{dietz_auditory_2011} and more complex models of auditory-based multimodal cognition \cite[e.g.,][]{Bustamante2018}. For this reason, combined with the increasing popularity of reproducible research~\cite{peng_reproducible_2011}, it is not surprising that there is an increasing number of auditory models available as software packages \cite[e.g.,][]{patterson_AIM_1995, fontaine_brian_2011,majdak_AMT_2021,AMT2021,Biberger2016, Biberger2016_code}.

However, models must be used with caution because they approximate auditory processes and are designed and evaluated under a specific configuration for a specific set of input sounds. While the evaluation conditions are selected to test the main properties of the simulated stages, models may provide different predictions when processing unseen sounds. Combined with the wide and low-threshold availability of model implementations, there is a chance of applying a model outside its specific signal or parameter range. Thus, studies comparing models' properties and configurations are important to model users. For example, Saremi \textsl{et al.} ~\cite{Saremi2016} compared seven models of cochlear filtering with respect to various parameters describing the nonlinear filtering process of an active cochlea, and Lopez-Poveda \cite{Lopez-Poveda2005} compared eight models of the auditory periphery based on the reproduction of auditory-nerve properties. Other related studies focus on a specific application \cite[e.g.,][]{anderson_comparison_1993, Breebaart2002, Harlander2014, Biberger2016, Biberger2016_code, Steinmetzger2019} or provide an introduction to modelling frameworks \cite{rudnicki_modeling_2015, dietz_framework_2018}. 

In the current study, we compare various monaural auditory models that approximate subcortical neural processing. For this comparison, we use model configurations that reduce the heterogeneity across model outputs, indicating advantages and disadvantages of these configuration choices. These configurations are evaluated using the same set of sound stimuli across models. The selected set of stimuli illustrates critical model properties that can be used as a guideline for the choice of a specific model. These properties include fast and slow temporal aspects, i.e., temporal fine structure and temporal envelope, that are evaluated for a wide range of presentation levels. 

We selected a number of auditory models that met two main criteria. First, the selected models describe the auditory path beginning with the acoustic input up to subcortical neural stages, in the cochlear nucleus (brainstem) and the inferior colliculus (midbrain). Consideration of these stages extends previous comparisons of auditory periphery models~\cite{Lopez-Poveda2005, Saremi2016}. Second, the model implementations are publicly available and validated to simulate psychoacoustic performance and/or physiological properties. We use the implementations available in the Auditory Modelling Toolbox (AMT) \cite{Soendergaard2013, majdak_AMT_2021, AMT2021}.

Based on our inclusion criteria, some models are excluded from the comparison, e.g., models that have only been evaluated at the level of cochlear filtering, such as models based on Hopf bifurcation~\cite{Kanders2017} and the model of asymmetric resonators with fast-acting compression~\cite{Lyon2011}. Other cochlear models, such as the Gammatone filter bank \cite{Hohmann2002}, the dual-resonance nonlinear model \cite[][]{Lopez-Poveda2001}, the chirp filter bank \cite{Tan2003} and the transmission-line model from \cite{Verhulst2012}, are included as modules in the selected models. We further excluded models whose structure did not contain one or more of the relevant processing stages that we evaluated in this study (Stages 3--6 in Fig.~\ref{fig:block-diagrams}). Two models that entered this category are the power spectrum models, EPSM \cite{Ewert2000} (no stages 3--5) and GPSM \cite{Biberger2016, Biberger2016_code} (no stage 5). Lastly, we did not include models focusing on specific psychoacoustic metrics \cite{Moore1997, Osses2016a, Taal2011}, despite the fact that such models are often based on comparable auditory stages as those described in this study. 

For the sake of simplicity, our analyses are focused on the comparison \textit{across} models rather than on a comparison with experimental data. Nevertheless, we provide experimental references to the simulations that are illustrated throughout this paper. Additionally, to encourage reproducible research in auditory modelling, all our paper figures can be retrieved using AMT~1.1, including (raw) waveform representations of intermediate model outputs.

The paper is structured as follows: In Sec.~\ref{sec:02-models} we provide a brief description of the processing stages in the selected auditory models. Their specific configurations are described in Sec.~\ref{sec:03-comparable}. The model comparison is presented in Sec.~\ref{sec:04-comparison-measures} and contains a description of the set of test stimuli as well as a detailed numerical description of the simulation results. Sec.~\ref{sec:05-perspective} starts with a list of applications of the evaluated models, including some general considerations for the application of auditory models in further modelling work. Note that although the detailed analysis in Sec. \ref{sec:04-comparison-measures} is relevant for model users who are interested in a transparent and accurate description of the illustrated model outputs, readers who are only interested in a bigger picture, as it is covered elsewhere \cite[e.g.,][]{Lopez-Poveda2005, Dau2008, meddis_computational_2010}, may wish to go directly to Secs. \ref{sec:05-perspective} and~\ref{sec:06-conclusions}.

\begin{table}[b]
    \centering
    \caption{List of selected models. The model labels used in this study correspond with the model functions in AMT~1.1.}\vspace{-12pt}
    \scalebox{0.85}{
    \begin{tabular}{lll} \hline\hline
         Label &  Reference \\ \hline
         \textsf{dau1997}      & Dau \textsl{et al.} (1997)~\cite{Dau1997b} \\         
         \textsf{zilany2014}   & Zilany \textsl{et al.} (2014)~\cite{Zilany2014} and Carney {et al.} (2015)~\cite{Carney2015}\\         
         \textsf{verhulst2015} & Verhulst \textsl{et al.} (2015)~\cite{Verhulst2015}\\         
         \textsf{verhulst2018} & Verhulst \textsl{et al.} (2018)~\cite{Verhulst2018a}\\
         \textsf{bruce2018}  & Bruce et al. (2018)~\cite{Bruce2018} and Carney \textit{et al.} (2015)~\cite{Carney2015}\\
         \textsf{king2019}   & King \textsl{et al.} (2019)~\cite{King2019} \\ 
         \textsf{relanoiborra2019} & Relaño-Iborra \textsl{et al.} (2019)~\cite{Relano-Iborra2019}\\
         \textsf{osses2021}  & Osses \& Kohlrausch (2021)~\cite{Osses2021a} \\          
         \hline\hline
    \end{tabular}
    }
    \label{tab:sel-models}
\end{table}

\section{Models}
\label{sec:02-models}

We define three model families, classified by their objectives~\cite{Gelfert2016}, which translate into three different levels of detail in simulating the cochlear processing, as schematised in Fig.~\ref{fig:model-families}. The selected models are listed in Table \ref{tab:sel-models} and are labelled throughout this paper by the last name of the first author and the year of the corresponding publication. This naming system directly reflects the corresponding model functions implemented in AMT~1.1~\cite{majdak_AMT_2021}. 

\begin{figure*} [ht!]
    \centering
    \includegraphics[width=0.98\textwidth]{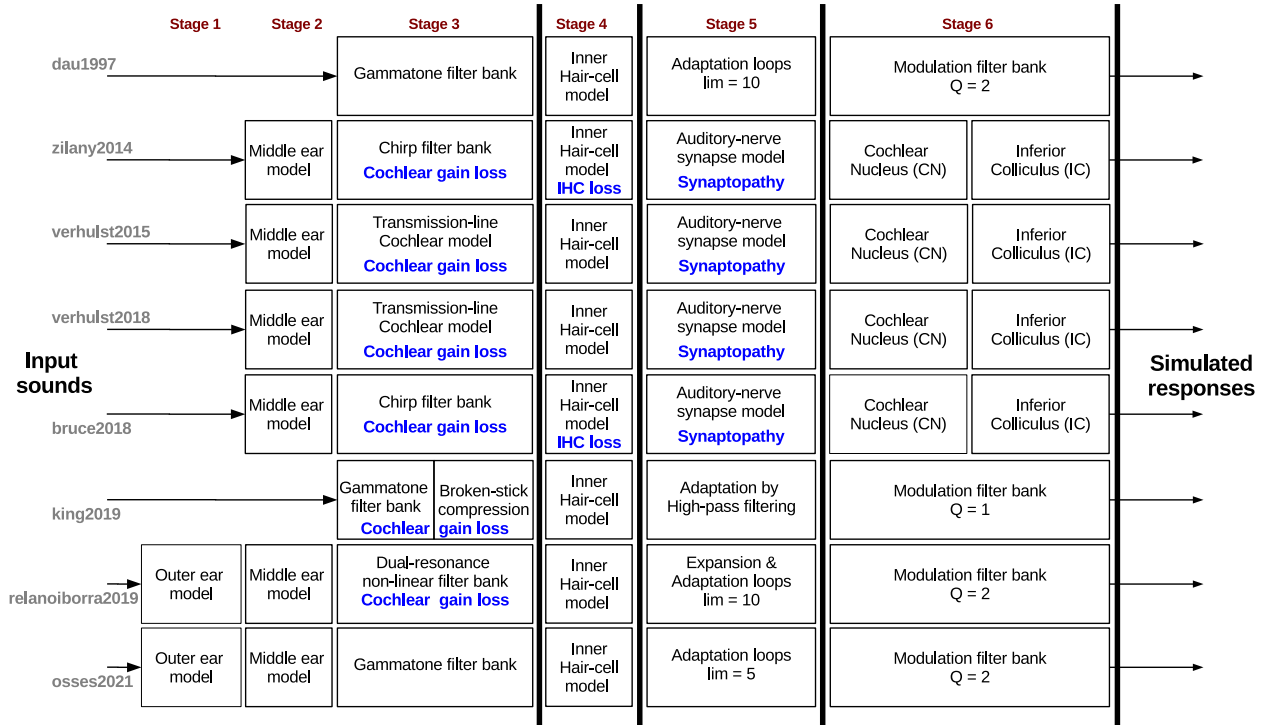}\\
    \vspace{-10pt}
    \caption{Block diagrams of the selected auditory models. Vertical lines: Intermediate model outputs as the basis for the evaluation in the corresponding sections. Blue: Type of hearing impairment that can be accounted for in the corresponding stage \protect(see a brief overview in Sec.~\ref{sec:discussion-features}).}
    \label{fig:block-diagrams}
\end{figure*}

We define the family of \textit{biophysical models} (Fig.~\ref{fig:model-families}\textbf{a}) that use a transmission line consisting of many resonant stages coupled by the cochlear fluid. Biophysical models aim at exploring how the properties of the system emerge from biological-level mechanisms, needing a fine-grained description at this level. The biophysical models are represented by \textsf{verhulst2015}~\cite{Verhulst2015} and its extended version, \textsf{verhulst2018}~\cite{Verhulst2018a} (model version 1.2 \cite{Osses2019d, Verhulst2020}). We further define \textit{phenomenological models} which primarily predict physiological properties of the system, using a more abstract level of detail than the biophysical models. The phenomenological models considered here rely on dynamically adapted bandpass-filtering stages combined with nonlinear mappings (Fig.~\ref{fig:model-families}\textbf{b}) and are represented by \textsf{zilany2014}~\cite{Zilany2014} and its extended version \textsf{bruce2018}~\cite{Bruce2018}, both combined with the same-frequency inhibition-excitation (SFIE) stages for subcortical processing~\cite{Nelson2004}. Further approximation is given by \textit{functional-effective models} \cite{Dau1996a}, which target the simulation of behavioural (perceptual) performance rather than the direct simulation of neural representations. These models usually approximate the cochlear processing by using static bandpass filtering with an optional nonlinear mapping (Fig.~\ref{fig:model-families}\textbf{c}). The linear effective models are represented by \textsf{dau1997}~\cite{Dau1997b} and \textsf{osses2021}~\cite{Osses2021a} and the nonlinear effective models are represented by \textsf{king2019}~\cite{King2019} and \textsf{relanoiborra2019}~\cite{Relano-Iborra2019}. Given that for each model a similar level of approximation has been generally used in the design of subsequent model stages, we use the defined categories to reflect the nature of the entire model. The defined categories are not meant to represent a hard boundary for model classification. Hence, we do not discard the existence of other more- or less-detailed models than the selected biophysical and effective models, respectively.

The selected monaural models share common stages of signal processing, as indicated in the schematic diagrams  of Fig.~\ref{fig:block-diagrams}, with some stages even using the same (digital) implementation. Each model stage mimics, with greater or lesser detail, underlying hearing processes along the ascending auditory pathway. The thick vertical lines in Fig.~\ref{fig:block-diagrams} indicate the intermediate model outputs which are the basis for our evaluation. Note that these stages are, conceptually speaking, independent of each other, however because of nonlinear interactions between them, processing performed by these stages is not commutative, thus requires a step-by-step approach. 

\subsection{Outer ear}

The listener's head, torso, and pinna filter incoming sounds. The ear-canal resonance further emphasises frequencies around 3000~Hz~\cite{Rosowski1991}. These effects can be accounted for by filtering the sound with a head-related transfer function (HRTF) \cite[e.g.,][]{moller1995} or by applying a headphone-to-tympanic-membrane transfer function, as used in \textsf{relanoiborra2019} and \textsf{osses2021}. The other six selected models that do not include an outer-ear filter, implicitly assume that either the outer ear does not introduce a significant effect in the subsequent sound processing chain, or that the sounds are presented near the tympanic membrane, as is the case for a sound presentation using in-ear earphones.


\subsection{Middle ear}
\label{sec:me}

\begin{figure}
    \centering
    \includegraphics[scale=0.5]{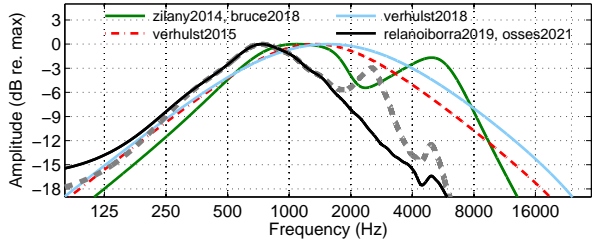}\\
  	\vspace{-10pt}
    \caption{Amplitude spectra of the four middle-ear filters used in six of the evaluated models. The lines were shifted vertically to display their individual maximum at 0\,dB. For \textsf{relanoiborra2019} and \textsf{osses2021}, the grey dashed line shows the combined response of the outer- and middle-ear filters. \textbf{Literature}: Fig.~1A from \cite{Puria2003} and Fig.~3 from~\cite{Pascal1998}.} %
    \label{fig:middleear}
\end{figure}

Six of the eight evaluated models include a stage of middle-ear filtering. The transfer functions of the middle-ear filters used in these models are shown in Fig.~\ref{fig:middleear}. 
The transfer functions in \textsf{verhulst2015} and \textsf{verhulst2018} approximate the middle-ear forward pressure gain (``M1'' in \cite{Puria2003}). The humanised \textsf{zilany2014} and \textsf{bruce2018} models use a linear middle-ear filter \cite{Ibrahim2010, Ibrahim2012} that approximates the forward-pressure measurements from~\cite{Puria1997, Pascal1998}. The middle-ear filters in \textsf{relanoiborra2019} and \textsf{osses2021} are those designed to represent stapes velocity near the oval window of the cochlea \cite{Lopez-Poveda2001, Goode1994}. These models use the same filter implementation, but the filter in \textsf{osses2021} includes a gain factor to provide a 0-dB amplitude in the frequency range of the passband and a fixed group-delay compensation.

Middle-ear filtering not only introduces a bandpass characteristic to the incoming signal (Fig.~\ref{fig:middleear}), but also affects the operating range of cochlear compression in models relying on nonlinear cochlear processing, i.e., \textsf{verhulst2015}, \textsf{verhulst2018}, \textsf{zilany2014}, \textsf{bruce2018}, and \textsf{relanoiborra2019}. The passband gains of the middle-ear filters are indicated in Table~\ref{tab:detailed-info} and range between $-66.9$\,dB (\textsf{relanoiborra2019}) and $+24$\,dB (\textsf{verhulst2015}). In nonlinear models, lower and higher passband gains vary the actual input level to the filter bank, shifting the onset of cochlear compression to higher and lower knee points, respectively. 

\subsection{Cochlear filtering}

A cochlear filter bank performs a spectral analysis of incoming signals to simulate the mechanical oscillations of the basilar membrane (BM) and organ of Corti at different points along the cochlea. Because of complex interactions between the BM, the cochlear fluid, and outer hair cells (OHCs), this analysis depends on the tonotopic position along the cochlea. All approaches used to simulate cochlear filtering produce a set of $N$ time-domain signals, for $N$ simulated characteristic frequencies (CFs). Each cochlear section is assumed to either have relatively sharp frequency tuning (Eq.\,\ref{eq:BW-Shera}, \cite{Shera2002}) or broader tuning (Eq.\,\ref{eq:BW-GM}, \cite{Glasberg1990}). For CFs expressed in Hz:

\vspace{-8pt}
\begin{eqnarray}
	\mbox{Q\textsubscript{ERB}} = 12.7 \cdot \left(\mbox{CF}/1000 \right)^{0.3}
	\label{eq:BW-Shera} \\
	\mbox{Q\textsubscript{ERB}}=\mbox{CF}/\left[ 24.7 \cdot \left(4.37\cdot \mbox{CF}/1000+1\right)\right]
	\label{eq:BW-GM}
\end{eqnarray}

The models \textsf{verhulst2015} and \textsf{verhulst2018} use a transmission-line model fitted to otoacoustic estimates of human cochlear filtering \cite{Verhulst2012}. In \textsf{zilany2014} and \textsf{bruce2018}, the filtering is based on a chirp filter bank \cite{Tan2003, Zilany2006} tuned to a human cochlea~\cite{Ibrahim2010, Ibrahim2012, ronne_modeling_2012}. The cochlear filters in these models are assumed to be tuned according to Eq.~\ref{eq:BW-Shera}.

In \textsf{dau1997} and \textsf{osses2021}, the linear Gammatone filter bank from \cite{Hohmann2002} is used. \textsf{King2019} uses the Gammatone filter bank from \cite{Hohmann2002} followed by a compressive stage acting above a given knee point. In \textsf{relanoiborra2019}, the cochlear processing is simulated by the dual-resonance nonlinear filter bank (DRNL) \cite{Lopez-Poveda2001}. The cochlear filters of these models are assumed to be tuned according to~Eq.~\ref{eq:BW-GM}.

\subsection{Inner hair cell}

The inner hair cells (IHCs) transform the mechanical BM and organ of Corti oscillations into receptor potentials, subsequently initiating neuronal discharges in the auditory nerve (AN)~\cite{Westerman1988}. In the most simple approach, the IHC processing can be simulated as an envelope extractor that removes phase information for high stimulus frequencies,  implemented as a half-wave rectification followed by a lowpass (LP) filter. This approach is used in \textsf{dau1997}, \textsf{king2019}, \textsf{relanoiborra2019}, and \textsf{osses2021}, in which the LP filters have $-3$-dB cut-off frequencies ($f$\textsubscript{cut-off}) between 1000 and 2000~Hz. In \textsf{zilany2014}, \textsf{bruce2018}, and \textsf{verhulst2015}, a nonlinear transformation is applied to the output of the cochlear filter bank, followed by a cascade of LP filters with $f$\textsubscript{cut-off} of 3000 Hz (\textsf{zilany2014} and \textsf{bruce2018}) and 1000~Hz (\textsf{verhulst2015}). The resulting $f$\textsubscript{cut-off} of each model ranges between 642~Hz (\textsf{verhulst2015}) and 1000~Hz (\textsf{dau1997, relanoiborra2019, king2019}), as indicated in Table~\ref{tab:detailed-info}. In \textsf{verhulst2018}, a more sophisticated IHC model is used \cite{Altoe2017}, that is implemented as a three-channel non-spiking Hodgkin-Huxley type model, with each of the channels representing mechanoelectrical and (fast and slow) potassium-gated processing~\cite{Altoe2017, Verhulst2018a}. 

\subsection{Auditory nerve}

The transduction from IHC receptor potentials into patterns of neural activity can be derived from the interaction between the IHC and AN. Several AN synapse models have been inspired by the three-store diffusion model \cite{Westerman1988}, assuming that the release of synaptic material is managed in three storage compartments. For steady-sound inputs, this model predicts a rapid neural firing shortly after the sound onset with a decreasing rate towards a plateau discharge rate, a phenomenon called adaptation \cite[e.g.,][]{Moore2013}. 

The AN synapse models in \textsf{verhulst2015}, \textsf{verhulst2018}, and \textsf{zilany2014} are based on \cite{Westerman1988}, but \textsf{zilany2014} further incorporates a power-law adaptation following the diffusion model from~\cite{Zilany2014}. The synapse model in \textsf{bruce2018} uses a diffusion model based on \cite{Peterson2018} to: (1) have limited release sites, and (2) come after the power-law adaptation instead of before it \cite{Bruce2018}. The outputs of these models simulate the firing of neurons\footnote{We adopted the word ``neuron,'' instead of the also-accepted spelling `neurone', following the etymological argument from \cite{Mehta2020}.} having a specific spontaneous rate of high-, medium-, and/or low-spontaneous rates.

The effective models, on the other hand, rely on a more coarse AN simulation, expressed in arbitrary units (a.u.). In \textsf{king2019}, adaptation is simulated by applying a  highpass filter with a cut-off frequency of 3\,Hz~\cite{King2019}. In \textsf{dau1997}, \textsf{relanoiborra2019}, and \textsf{osses2021}, adaptation is simulated by so-called adaptation loops \cite{Dau1996a} that introduce a nearly logarithmic compression to stationary input signals and a linear transformation for fast signal fluctuations (Appendix~B in \cite{Osses2021a}). The arbitrary units of these transformed outputs are named model units (MUs).

\subsection{Subcortical neural processing}
\label{sec:subcortical}
AN firing patterns propagate to higher stages along the auditory pathway, first through the auditory brainstem, then  towards more cortical regions \cite{Majdak2020}. On its way, AN spiking is mapped onto fluctuation patterns by neurons that are sensitive to the amplitude of low-frequency fluctuations \cite{Carney2018a}. This fluctuation sensitivity has been approximated using various approaches. Our analyses focus on model approximations of the modulation processing circuits of the ventral cochlear nucleus (CN) and inferior colliculus (IC) \cite{Nelson2004}, as well as on different modulation-filter-bank variants \cite{Dau1997b, Ewert2000}. As a result, we exclude the analysis of other subcortical structures such as those that play a particular role in the binaural interaction between ears (e.g., the dorsal cochlear nucleus and superior olive) \cite{Ashida2017, Majdak2020}. 

The modulation processing in the ventral CN and IC can be simulated using the same-frequency inhibition-excitation (SFIE) model, resulting in a widely tuned modulation filter (Q factor$\approx$1) with a best-modulation frequency (BMF) depending on the parameters of the model~\cite{Nelson2004, Carney2015}. The SFIE model has already been used in combination with the biophysical and phenomenological models described here. For example, \textsf{zilany2014} has been combined with the SFIE model using between one and three modulation filters \cite[e.g.,][]{Maxwell2020}. Or, \textsf{verhulst2015} and \textsf{verhulst2018} have used the SFIE model with one modulation filter centred at a BMF of 82.4~Hz (see Table~\ref{tab:detailed-info}) \cite{Verhulst2018a, Osses2019d}. Further, \textsf{bruce2018} can be combined with the SFIE model in the UR EAR 2020b toolbox \cite{Carney2020a}. Note that \textsf{zilany2014}, \textsf{verhulst2015}, and \textsf{verhulst2018} have used the output of their mean firing rate generator---an output that can be conceptualised as peri-stimulus time histograms (PSTHs) \cite{Gerstner2014}---as an input to the SFIE model. In \textsf{bruce2018}, because of the stochastic processes in its spike generator, repeated processing of the same stimulus is recommended to obtain a faithful PSTH that can appropriately account for power-law adaptation properties (see Sec.~3 in \cite{Bruce2018}).

The effective models, on the other hand, approximate the subcortical neural processing based on the modulation-filter-bank concept~\cite{Dau1997b, Ewert2000}. In \textsf{dau1997}, \textsf{king2019}, \textsf{relanoiborra2019}, and \textsf{osses2021}, linear modulation filter banks are used, covering a range of BMFs up to 1000~Hz. In \textsf{dau1997}, twelve modulation filters with a Q-factor of 2 and overlapped at their $-3$~dB points are used. The same modulation filters are used in \textsf{relanoiborra2019} and \textsf{osses2021}, but an additional 150-Hz LP filter is applied~\cite{Kohlrausch2000, Ewert2000} and the number of filters is limited so that the highest BMF is less than a quarter of the corresponding CF~\cite{Verhey1999}. In \textsf{king2019}, the filter bank is used with a wider tuning (Q=1, as suggested in \cite{Ewert2000,Ewert2002}), using ten 50\%-overlapped filters having a maximum BMF of 120~Hz~\cite{King2019}.

\section{Model configuration} 
\label{sec:03-comparable}

We evaluated the intermediate model outputs that are indicated by thick vertical black lines in Fig.~\ref{fig:block-diagrams}. The evaluation points are located after the cochlear filter bank (Stage 3), the IHC processing stage (Stage 4), the AN synapse stage or equivalent (Stage 5), and after the IC processing stage or equivalent (Stage 6). Starting with the default parameters of each model, we introduced small adjustments to obtain the most comparable model outputs. All comparisons can be reproduced with the function \textsf{exp\_osses2022} from AMT 1.1~\cite{majdak_AMT_2021}.

\subsection{Level scaling}

The same set of sound stimuli was used as input to all models. The waveform amplitudes were assumed to represent sound pressure expressed in Pascals (Pa). The models  \textsf{zilany2014}, \textsf{verhulst2015}, \textsf{verhulst2018}, \textsf{bruce2018}, and \textsf{relanoiborra2019} use this level convention and did not require further level scaling. The models \textsf{dau1997}, \textsf{king2019}, and \textsf{osses2021} interpret sound pressures between $-1$ and $1$\,Pa as amplitudes in the range $\pm0.5$, thus a factor of 0.5 (attenuation by 6\,dB) was applied to the generated stimuli to meet the level convention of these models. For these latter models, which include mostly level-independent stages, such calibration is relevant because the adaptation loops (used in \textsf{dau1997}, \textsf{osses2021}, also extensible to \textsf{relanoiborra2019}) include level-dependent scaling (Eqs. B1--B3 in \cite{Osses2021a}). In \textsf{king2019}, a calibrated knee point (default of 30\,dB) is used in its cochlear compression stage (Stage~3). All signal levels are reported as root-mean-square (rms) values referenced to 20\,$\mu$Pa, in dB sound pressure level (dB~SPL).

\subsection{Cochlear filtering}

The phenomenological and effective models can be set to simulate responses at any CF. However, because of the nature of the transmission-line structure, the selected biophysical models have a discrete tonotopy that translates into a discrete set of available CFs. 

The models \textsf{verhulst2015} and \textsf{verhulst2018} were set to 401 cochlear sections spaced at $\Delta x=$0.068\,mm with tonotopic distances $x_n$ ranging between $x_1=3.74$\,mm and $x_{401}=30.9$\,mm, that are related to CFs between CF$_1=12010$\,Hz and CF$_{401}=113$\,Hz, according to the apex-to-base mapping~\cite{Greenwood1990},  

\begin{equation}
    \mbox{CF}_n=A_0\cdot \left( 10^{-a \cdot x_n/1000} \right) - A \cdot k
    \label{eq:greenwood}
\end{equation}

\noindent where $x_n$ (in mm) can be obtained as $x_1+\Delta x\cdot\left(n-1\right)$, and $A=165.4188$\,Hz, $a=61.765$~1/m, $k=0.85$, and $A_0=20682$\,Hz. Note that when reporting results, we indicate the cochlear section number $n$ and its corresponding CF$_n$.

The cochlear-filtering parameters of \textsf{zilany2014} and \textsf{bruce2018} were those adapted to a human cochlea \cite{Ibrahim2010, Ibrahim2012}. Moreover, in order to analyse separately the effects of cochlear filtering and IHC processing in \textsf{zilany2014}, the outputs from the chirp filters representing the static and OHC-controlled filters (C2 and C1 in \cite{Zilany2014}) were added and analysed before the IHC nonlinear mapping was applied. This analysis follows a similar rationale as analysing the main output of the DRNL filter bank in \textsf{relanoiborra2019} (see Fig.~3a from \cite{Lopez-Poveda2001}). 

Finally, in \textsf{king2019}, we used a compression factor of 0.3 for all simulated CFs, which is different from the one-channel (on-CF) compression used in~\cite{King2019, Wallaert2018}.

\subsection{Inner hair cell and auditory nerve}

Default parameters were used for the IHC and AN stages of the evaluated effective models. However, the biophysical and phenomenological models require the choice of parameters to simulate a population of AN fibres. For each CF we simulated 20 fibres, having either high- (HSR), medium- (MSR), or low-spontaneous rates (LSR), distributed in a 0.6-0.2-0.2 ratio~\cite{Liberman1978, Liberman1990}, resulting in a 12-4-4 configuration (HSR-MSR-LSR). Note that for \textsf{verhulst2015} and \textsf{verhulst2018}, this deviates from the standard 13-3-3 configuration \cite{Verhulst2015, Verhulst2018a}. For \textsf{verhulst2015} and \textsf{verhulst2018}, the spontaneous rates of each fibre type were 68.5, 10, and 1 spikes/s for HSR, MSR, and LSR, respectively, as used in human-tuned simulations \cite[][]{Verhulst2018a}. For \textsf{zilany2014}, the spontaneous rates of each fibre type were 100, 4, and 0.1 spikes/s and for \textsf{bruce2018} were 70, 4, and 0.1 spikes/s for HSR, MSR, and LSR, respectively. We further disabled the random fractional noise generators in \textsf{zilany2014} and \textsf{bruce2018} \cite{Zilany2009}, and the random spontaneous rates in \textsf{bruce2018} (``std'' from Tab.~I in~\cite{Bruce2018} was set to zero). With this configuration,  the mean-rate synapse outputs of \textsf{verhulst2015}, \textsf{verhulst2018}, \textsf{zilany2014}, and \textsf{bruce2018} are deterministic. For this reason, to obtain population responses, we simulated the AN processing of each type of neuron only once and then weighted them by factors of 0.6, 0.2, and 0.2 for HSR, MSR, and LSR fibres, respectively. In contrast, the PSTH outputs that are reported for \textsf{zilany2014} and \textsf{bruce2018} are not deterministic, requiring the simulation of each AN fibre for each CF. Therefore, PSTH population responses were obtained by counting the average number of spikes in time windows of 0.5~ms across 100 repetitions of the corresponding stimuli.

\subsection{Subcortical neural processing} 

The default configuration of the model stages of subcortical processing (Stage 6, Fig.~\ref{fig:block-diagrams}) differs in the number of modulation filters (from 1 to 12) and in their tuning across  models. In our study, we use only one modulation filter targeting a BMF of approximately 80~Hz (see ``Theoretical BMF'' in Table~\ref{tab:detailed-info}) and a Q-factor of approximately~1 for \textsf{zilany2014}, \textsf{verhulst2015}, \textsf{verhulst2018}, \textsf{bruce2018}, and \textsf{king2019}, and a Q-factor of 2 for \textsf{dau1997}, \textsf{relanoiborra2019}, and \textsf{osses2021}.

For the biophysical and phenomenological models, we used the SFIE model~\cite{Nelson2004, Carney2015} using two different configurations. The SFIE model \cite{Nelson2004} integrated in \textsf{verhulst2015} and \textsf{verhulst2018} has CN parameters with excitatory and inhibitory time constants of $\tau\textsubscript{exc}=0.5$\,ms and $\tau\textsubscript{inh}=2$\,ms, a delay $D=1$\,ms, and a strength of inhibition of $S=0.6$. The IC stage uses $\tau\textsubscript{exc}=0.5$\,ms, $\tau\textsubscript{inh}=2$\,ms \cite{Verhulst2018a}, $D=2$\,ms, and $S=1.5$ \cite{Nelson2004}, achieving a BMF of 82.4\,Hz. 

For \textsf{zilany2014} and \textsf{bruce2018}, the SFIE model is a separate stage \cite{Carney2015}, implemented as \textsf{carney2015} in AMT~1.1, where either the mean-rate (\textsf{zilany2014}) or the PSTH outputs (\textsf{bruce2018}) are used as inputs. In our analysis, we only used the output of the band-enhanced IC cell, which corresponds to the SFIE model from \cite{Nelson2004}. The CN parameters were identical to those for the biophysical models. The IC parameters were $\tau\textsubscript{exc}=1.11$~ms, $\tau\textsubscript{inh}=1.67$~ms, $D=1.1$~ms, and $S=0.9$, achieving a BMF of 83.9~Hz \cite{Carney2015}. Note the different inhibition strength $S$ between models. In the biophysical models, the IC output is dominated by inhibitory responses ($S>1$) whereas in the phenomenological models the IC output is dominated by excitatory responses ($S<1$). 

\section{Evaluation}
\label{sec:04-comparison-measures}

\begin{figure*}
    \centering
    \includegraphics[width=\textwidth,clip=true]{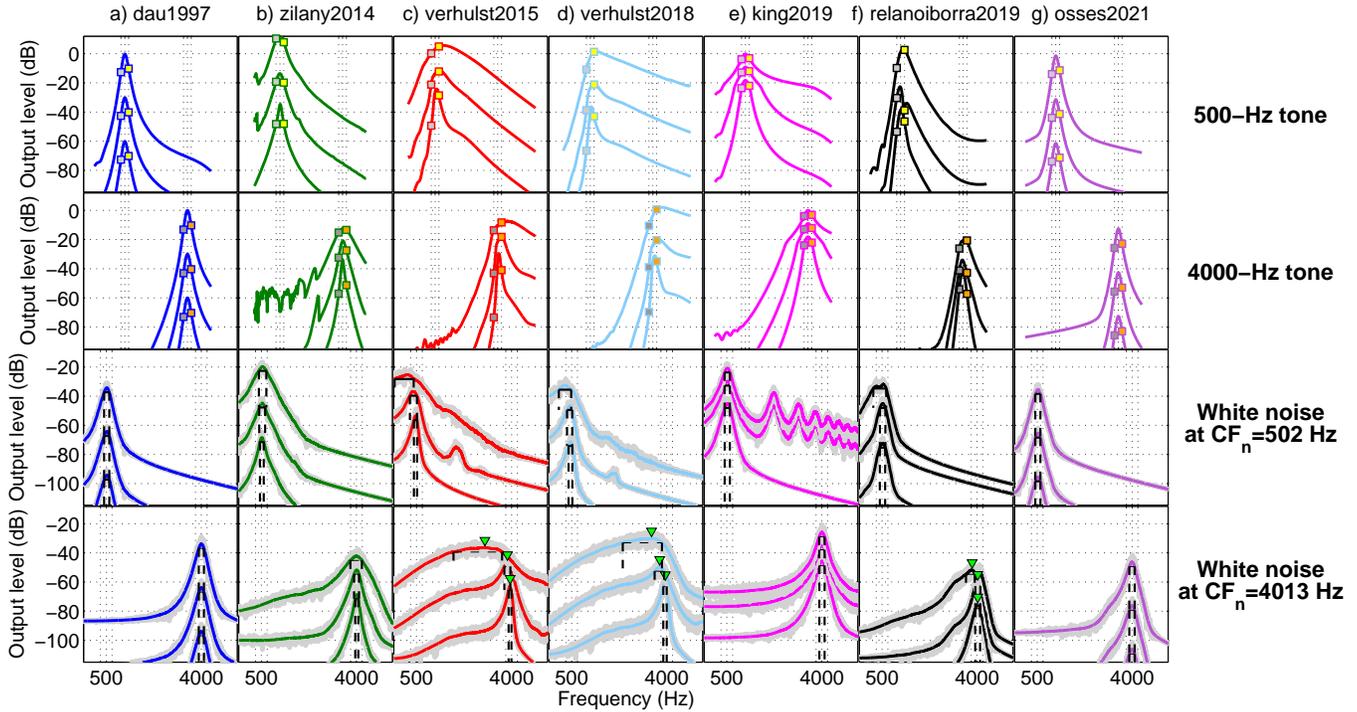}\\
    \vspace{-12pt}
    \caption{Filter bank responses to pure tones at 500\,Hz and 4000\,Hz (top two rows) and spectral magnitudes of single-channel responses to white noises (bottom two rows) for sounds of 40-, 70-, or 100-dB SPL (bottom-to-top coloured curves, respectively). 
    In the first two rows, the responses represent excitation patterns at the simulated CFs. The coloured markers indicate the amplitudes at the (off-frequency) CFs one ERB$_N$ below (grey) and one ERB$_N$ above (yellow or orange) the on-frequency CF (502 or 4013 Hz). These markers are highlighted using the same colours in Fig.\,\ref{fig:IO-filterbank}. In the third and fourth rows, the average FFT response of two cochlear-filters (CFs of 502 Hz or 4013 Hz) in response to six 500-ms white noise sections are shown in grey, and the corresponding smoothed responses are shown in colour. This type of responses was used to assess the quality factors of Fig.~\ref{fig:noise-filterbank}. The dashed black lines indicate the corresponding $-3$-dB filter bandwidths. All responses were shifted vertically by the reference gains given in Table~\ref{tab:detailed-info} (see the text for details). \textbf{Literature}: Fig.~1A from \cite{Ren2002}, Fig. 2C--E from \cite{Jepsen2008}, and Fig.~2 from~\cite{Saremi2016}.}
    \label{fig:Excitation}
\end{figure*}

\begin{figure*}
    \centering
    \includegraphics[scale=0.49,clip=true]{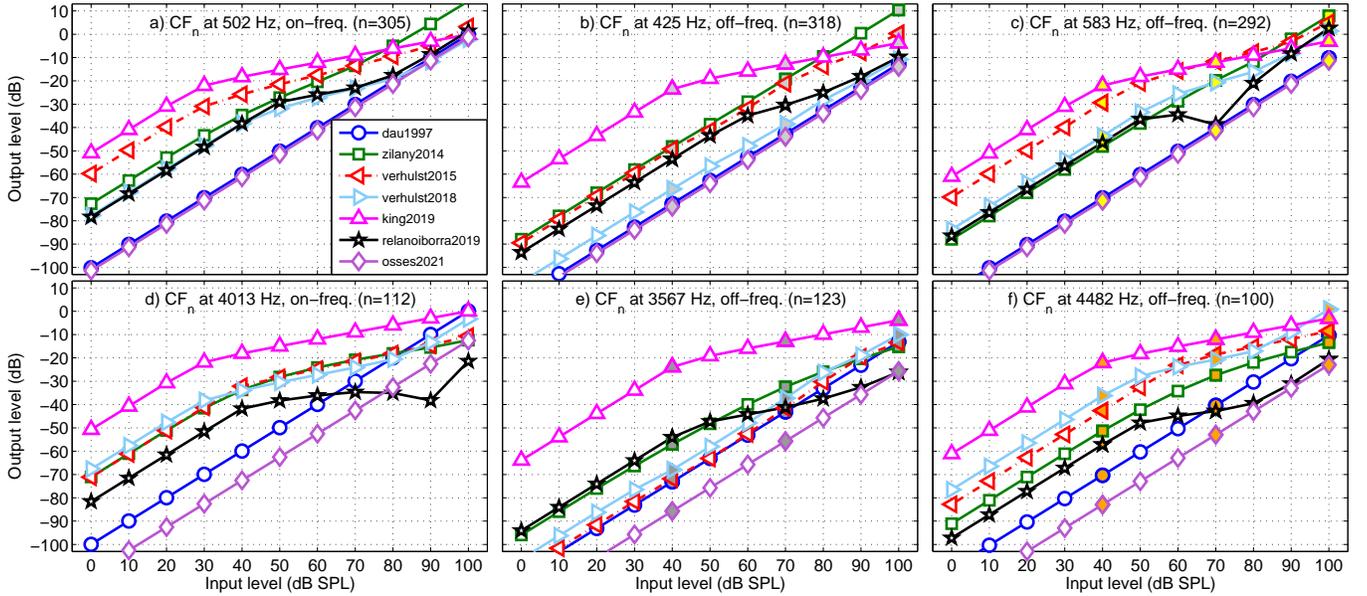}\\
    \vspace{-10pt}
    \caption{Input-output (I/O) curves for pure tones at 500\,Hz (panels \textbf{a--c}) and 4000\,Hz (panels \textbf{d--f}), at six model CF$_n$ frequencies (see Eq.~\ref{eq:greenwood}). \textbf{Left (a,d)}: On-frequency simulations, i.e., output of the cochlear filter with the CF tuned to that of the stimulus frequency. \textbf{Middle (b,e), right (c,f)}: Off-frequency simulations, one ERB below and above the on-frequency, respectively. The exact simulated on- and off-frequency CFs are indicated in the title of each panel. The filled markers indicate off-CF amplitudes that are also highlighted in the corresponding frequency responses of Fig.~\ref{fig:Excitation}. All I/O curves were shifted vertically by the reference gains given in Table~\ref{tab:detailed-info} (see the text for details). \textbf{Literature}: Figs.~1--3 from \cite{Robles2001} and Fig.~3 from~\cite{Saremi2016}.}
    \label{fig:IO-filterbank}
\end{figure*}

In this section we analyse the outputs of the eight selected auditory models in a number of test conditions, whose results are presented in Figs.~\ref{fig:Excitation}--\ref{fig:mfb-click}. We aimed at a comparison across models and thus, for the sake of clarity, we refrained from a direct comparison to ground-truth references from physiological data. However, such a comparison is important and interesting. For this reason, we provide references where similar experimental and/or simulation analyses have been presented. These references are indicated as ``Literature'' in the caption of the corresponding figure. Alternatively, the outputs of the biophysical and phenomenological models may be considered as referential because they have been primarily developed to reflect physiological (human or animal) responses to sounds. This latter assumption always requires a careful consideration, especially when translating findings from animal to human physiology.

\subsection{Cochlear filtering}

Sound processing in the cochlea depends not only on the frequency but also on the level of the input stimulus \cite{Recio2000}. The amplitude of the BM vibration displacement increases for higher levels, following an amplitude growth that comprises linear and compressive regimes \cite[][]{Robles2001}. We illustrate this level dependency for a set of pure tones and white noises. The pure tones had frequencies of 500 or 4000\,Hz, with a duration of 100 ms. The white noise was generated as a fixed 3-s long waveform 
 with a flat spectrum between 20 and 20000 Hz. All sounds were gated on and off with a 10-ms raised-cosine ramp and had levels of 40, 70, and 100\,dB\,SPL. The obtained responses are shown in Fig.~\ref{fig:Excitation}, which were vertically shifted by the gains indicated in Table~\ref{tab:detailed-info}. These gains were derived for each model using the 1000-Hz pure tone of 100\,dB SPL as a reference. The frequency responses were level-dependent for \textsf{zilany2014}, \textsf{verhulst2015}, \textsf{verhulst2018}, \textsf{king2019}, and \textsf{relanoiborra2019}. For  \textsf{verhulst2015}, \textsf{verhulst2018}, and \textsf{relanoiborra2019}, we further observed a change in the location of their maximum amplitude (green triangles in Fig.~\ref{fig:Excitation}, fourth row). Although a shift of the responses with increasing the stimulus level is supported by experimental data \cite{McFadden1983, Recio2000}, this argument was later challenged \cite{Moore2003a} and rather attributed to shallower upper tails in the responses. A final observation is that \textsf{king2019}, \textsf{zilany2014}, \textsf{verhulst2015} and \textsf{verhulst2018}, showed a certain amount of distortion in the frequency responses of 70- and 100-dB SPL sounds (Fig.~\ref{fig:Excitation}, second and third rows). For the responses to white noises at CF=502~Hz, the distortions occur in the upper tail of the cochlear responses as a side effect of the (amplitude) compression stage. This frequency-response distortion is most prominent in \textsf{king2019} due to its broken-stick compression (rise of the amplitudes to the power of 0.3).\footnote{Note that \textsf{king2019} was primarily developed to simulate responses to pure tones or narrow-band signals using cochlear channels with CFs that are either on-frequency (one cochlear channel) \cite{Wallaert2017} or spanning $\pm$2\,ERB$_N$ around the on-frequency CF (five cochlear channels) \cite{King2019, Wallaert2018}. The prominent frequency distortions shown in Fig.~\ref{fig:Excitation}\textbf{e} (third row) are thus outside the tested CF ranges for this model.}
 
 
 


\subsubsection{Compressive growth}

The set of pure tones was further used to assess the curves relating the input stimulus levels with levels at the output of the cochlear filter banks, known as input-output (I/O) curves.
For this analysis we included stimulus levels between 0 and 100 dB SPL in steps of 10 dB. The I/O curves were obtained for (1) the on-frequency CF tuned to the frequency of the input stimulus, and (2) the off-frequency responses of cochlear filters tuned to one equivalent rectangular bandwidth number (ERB$_N$) \cite[][]{Glasberg1990} below and above the stimulus frequency.

The obtained I/O curves are shown in Fig.~\ref{fig:IO-filterbank} for on-frequency (left panels) and off-frequency simulations ($\pm$1\,ERB$_N$, middle and right panels). The I/O curves were vertically shifted by the reference gains indicated in Table~\ref{tab:detailed-info} (as in Fig.~\ref{fig:Excitation}). As expected for the level-independent Gammatone filters used in \textsf{dau1997} and \textsf{osses2021}, the curves were linear in all panels of Fig.~\ref{fig:IO-filterbank}. For the remaining models, more compressive behaviour was observed for on-frequency curves (left panels) while more linear curves were obtained for off-frequency CFs (middle and right panels), except for \textsf{relanoiborra2019} and \textsf{king2019}, that had on- and off-frequency compression. 

For \textsf{zilany2014}/\textsf{bruce2018}, the I/O curves were fairly linear in response to 500-Hz tones (top panels) for both on- and off-frequency CFs. For 4000-Hz tones, a prominent compressive behaviour was observed in the on-frequency curves (Fig.~\ref{fig:IO-filterbank}\textbf{d}) where, additionally, the curve for \textsf{verhulst2018} turned from a compressive to a linear regime for signal levels above 80\,dB. The off-frequency I/O curves obtained for \textsf{verhulst2018} were similar to those for \textsf{verhulst2015} but had overall lower and higher amplitudes for the pure tones of 500\,Hz (Fig.~\ref{fig:IO-filterbank}\textbf{b--c}) and 4000\,Hz (Fig.~\ref{fig:IO-filterbank}\textbf{e--f}), respectively, as a consequence of the differences in their middle-ear filters (see Fig.~\ref{fig:middleear}). The tendency to a more linear regime in off-frequency CFs has been shown previously~\cite{Robles2001}. This is in fact the basis for having compression only applied to the on-frequency channel in \textsf{king2019}~\cite{King2019, Wallaert2018}. However, the default compression rate of 0.3 for the on-frequency channel with no compression for off-frequency channels leads to an unrealistic level balance between on- and off-frequency channels.

\subsubsection{Frequency selectivity: Filter tuning}
\label{sec:freq-sel}

The frequency selectivity of each filter bank was computed in response to the described frozen noise waveform, presented at 40, 70, and 100\,dB\,SPL. The estimates of frequency selectivity were obtained from FFT responses averaged across 500-ms non-overlapped analysis windows, meaning that the estimates were obtained from six statistically-independent noise sections.

The frequency response of thirty-two filters with CFs between 126\,Hz ($n$=396 in Eq.\,\ref{eq:greenwood}) and 9587\,Hz ($n$=24 in Eq.\,\ref{eq:greenwood}) at steps of $n$=12 bins was obtained. For illustration purposes, we also included in this analysis the on-frequency CFs used in Fig.~\ref{fig:IO-filterbank} (CF=502 Hz, $n=305$; CF=4013 Hz, $n=112$), whose obtained responses are shown in Fig.~\ref{fig:Excitation}. For each filter response, a quality factor Q\textsubscript{$-$3 dB}=CF/BW was obtained, where BW is the bandwidth defined by the lower and upper 3-dB down points of each filter transfer function (black dashed lines in Fig.~\ref{fig:Excitation}).

\begin{figure}
	\centering
	\includegraphics[scale=0.48]{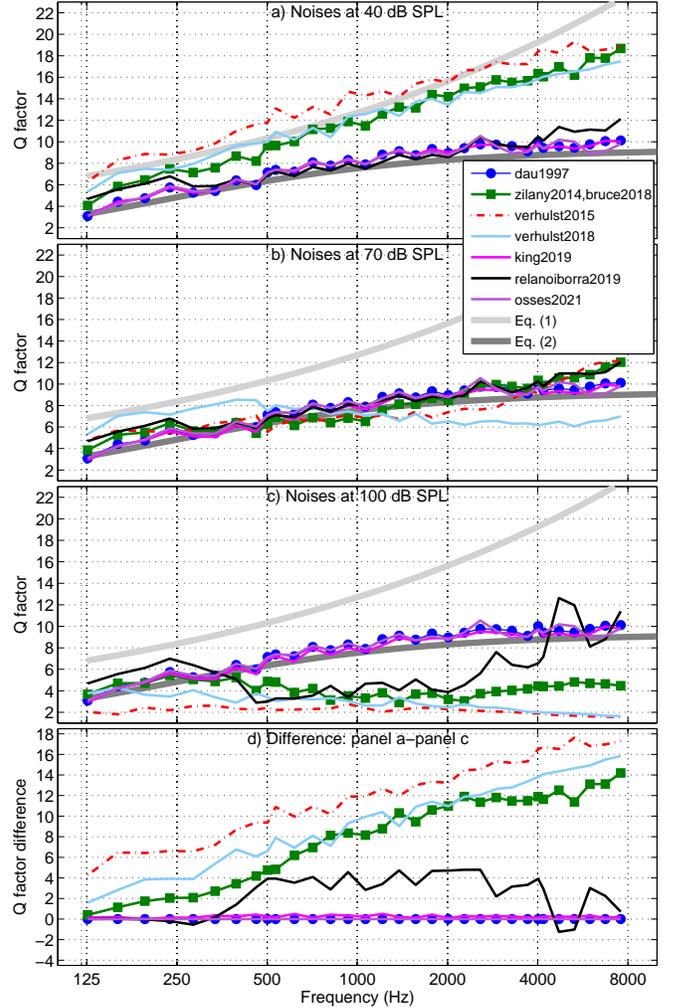}\\
	\vspace{-10pt}
	\caption{Filter tuning expressed as quality factors Q for noises of 40, 70, and 100\,dB SPL (panels \textbf{a--c}), and Q-factor difference obtained from the results of 40- and 100-dB noises (panel~\textbf{d}). \textbf{Literature}: Fig.~4 from \cite{Shera2002} and Fig.~4B from~\cite{Saremi2016}.}	
	\label{fig:noise-filterbank}
\end{figure}

The frequency-selectivity simulations for each of the filter banks are shown in Fig.~\ref{fig:noise-filterbank} for noises of 40 (panel\,\textbf{a}), 70 (panel\,\textbf{b}), and 100\,dB~SPL (panel\,\textbf{c}). The analytical filter tuning curves given by Eqs.\,\ref{eq:BW-Shera} and \ref{eq:BW-GM} are indicated as light and dark grey traces in Fig.~\ref{fig:noise-filterbank}. Note that with this comparison, we assume that the Q factors within one ERB are similar to Q\textsubscript{$-3$ dB} values. The results for 40-dB noises show that the frequency selectivity follows either the analytical tuning of Eq.\,\ref{eq:BW-Shera} (\textsf{zilany2014}, \textsf{bruce2018}, \textsf{verhust2015}, and \textsf{verhulst2018}) or the tuning of Eq.\,\ref{eq:BW-GM} (\textsf{dau1997}, \textsf{relanoiborra2019}, \textsf{king2019}, and \textsf{osses2021}). When looking at the results for higher levels (Fig.~\ref{fig:noise-filterbank}\textbf{b--c}), no change in tuning was observed for \textsf{dau1997} and \textsf{osses2021}, as expected for linear models. For the nonlinear models, the results for 70-dB noises in Fig.~\ref{fig:noise-filterbank}\textbf{b} showed overall lower Q factors, but with only a small change for \textsf{king2019} and \textsf{relanoiborra2019}. The results for 100-dB noises in Fig.~\ref{fig:noise-filterbank}\textbf{c} showed a further lowering of Q factors in the biophysical and phenomenological models, reaching values as low as Q $\approx2$ in \textsf{verhulst2015}, lower Q factors for frequencies up to about 4000~Hz in \textsf{relanoiborra2019}, and virtually unaffected Q factors in \textsf{king2019}. A closer inspection to the outputs of \textsf{king2019} revealed that there was a filter broadening as a consequence of its broken-stick nonlinearity stage, but this broadening predominantly affected the frequency responses outside the range defined by the 3-dB bandwidth used to derive the Q factors (see Fig.~\ref{fig:Excitation}\textbf{e}). To illustrate the Q-factor transition when increasing the signal level in each model, the difference between Q factors obtained from 40- and 100-dB noises is shown in Fig.~\ref{fig:noise-filterbank}\textbf{d}, where a decrease in Q factor with increasing signal level is represented by a positive Q-factor difference. 

Additionally, we observed that \textsf{relanoiborra2019} and \textsf{king2019} introduce a change in selectivity at overall higher levels compared to the biophysical and phenomenological models. A closer look at this aspect revealed that this change occurs because \textsf{relanoiborra2019} and \textsf{king2019} only apply compression after the bandpass filtering and, therefore, lower level signals are used as input for their compression (broken-stick) module. 

\begin{figure*}
    \centering
	\includegraphics[scale=0.6,clip=true,trim=0 0 0 0cm]{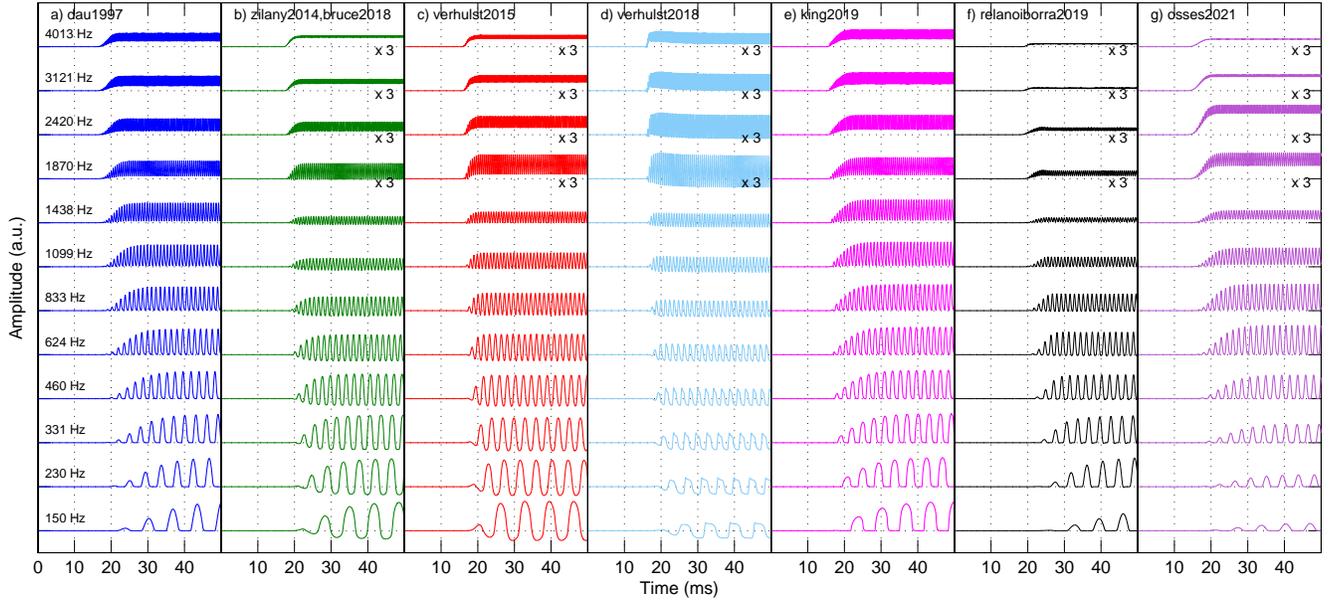}\\
	\vspace{-10pt}
	\caption{Simulated IHC responses to pure tones of different frequencies evaluated at the corresponding on-frequency bin. The amplitudes were normalised with respect to their maximum value to allow a direct comparison across models. \textbf{Literature}:~Fig.~9 from \cite{Palmer1986} and Fig.~7 from \cite{Lopez-Poveda2006}.}
    \label{fig:ihc-wave}
\end{figure*}

\subsubsection{Frequency selectivity: Number of filters}

The number of filters in a filter bank is relevant for several model applications because too few filters can lead to a loss of signal information \cite[e.g.,][]{Antoni2010} and too many filters may unnecessarily increase the computational costs. The number of filters is a free parameter in \textsf{zilany2014}/\textsf{bruce2018}, but is fixed for \textsf{verhulst2015} and \textsf{verhulst2018} to yield an accurate precision of the transmission-line solver~\cite{Altoe2014}. The remaining models use by default one ERB-wide bands (\textsf{dau1997}, \textsf{king2019}, and \textsf{osses2021}), or have an overlap every 0.5~ERB$_N$ (\textsf{relanoiborra2019}). 

Here, we report the minimum number of filters that are required to obtain a filter bank with overlapping at $-3$-dB points of the individual filter responses. Using the empirical Q-factors of Fig.~\ref{fig:noise-filterbank}, we assessed the number of filters that would be required to cover a frequency range between 126\,Hz ($n=396$ in Eq.~\ref{eq:greenwood}) and the first filter with its upper cut-off frequency equal or greater than 8000\,Hz. The number of filters derived from the 40-dB and 100-dB frequency tuning curves (Fig.~\ref{fig:noise-filterbank}\textbf{a,c}) are shown in Table~\ref{tab:detailed-info}, including the average filter bandwidth in ERB$_N$ for the corresponding model.

For the biophysical models, the filters were much wider at the higher level than for the other models, with average bandwidths being as wide as 3.05\,ERB$_N$ for \textsf{verhulst2015} and 2.30\,ERB$_N$ for \textsf{verhulst2018}. This contrasts with the 1.57\,ERB$_N$ for \textsf{zilany2014} and \textsf{bruce2018} and the 1.15\,ERB$_N$ or less for the remaining models. These bandwidths are a consequence of the fast-acting (sample-by-sample) compression that is applied just before the transmission-line in the biophysical models and the slower-acting bandwidth control in \textsf{zilany2014} (denoted as the ``control path'' in the chirp-filter bank). While cochlear filters are generally wider at high sound levels~\cite[e.g.,][]{Ruggero1997, Ren2002}, the appropriate tuning must be evaluated depending on  the species' characteristics, the tested CFs, and the type of evaluated excitation signals. 

\subsection{IHC processing: Phase locking to temporal fine structure}
\label{sec:IHC-proc}

To illustrate the loss in phase locking to temporal fine structure with increasing  stimulus frequency, we simulated IHC responses to pure tones with frequencies between 150\,Hz ($n$=387 in Eq.\,\ref{eq:greenwood}) and 4013\,Hz ($n$=112 in Eq.\,\ref{eq:greenwood}) spaced at $n$=25 bins, resulting in twelve test frequencies. The tones were generated at 80\,dB SPL, with a duration of 100\,ms, and were gated on and off with 5-ms raised-cosine ramps. The simulated waveforms, that are assumed to approximate the IHC potential, are displayed and described in terms of AC (fast-varying) and DC (average bias) components, and the simulated resting potentials ($V\textsubscript{rest}$). The AC potential was assessed from the peak-to-peak amplitudes as $V\textsubscript{AC}=V\textsubscript{peak,max}-V\textsubscript{peak,min}$. The DC potential was obtained as $V\textsubscript{DC}=\left(V\textsubscript{peak,max}+V\textsubscript{peak,min}\right)/2-V\textsubscript{rest}$~\cite{Lopez-Poveda2006, Palmer1986}.

The obtained IHC waveforms are shown in Fig.~\ref{fig:ihc-wave}. Within each panel, bottom to top waveforms represent on-frequency simulations for the test signals, from low to high frequency carriers, respectively. For some model outputs, the four highest carriers (1870$\leq f_c\leq$4013~Hz) were amplified by a factor of 3 to improve waveform visibility. The simulated voltages before the tone onset, i.e., the resting potential $V\textsubscript{rest}$, were equal to 0 for all models except for \textsf{verhulst2018}, where $V\textsubscript{rest}$ was $-57.7$\,mV (not schematised in Fig.~\ref{fig:ihc-wave}). It seems clear, however, that the decrease of peak-to-peak AC voltage towards high frequencies---a measure of the residual amount of temporal fine structure---is significantly different across models. When increasing the CFs from 1099 to 4013\,Hz, three models showed $V\textsubscript{AC}$ reductions of less than 76.0\% (\textsf{king2019}: 59.7\%-decrease from $3.12\cdot10^{-3}$ to $1.26\cdot10^{-3}$\,a.u.; \textsf{dau1997}: 62.6\%-decrease from 0.097 to 0.037\,a.u.; and \textsf{verhulst2018}: 76.0\%-decrease from 39.2 to 9.4\,mV), while the other five models showed $V\textsubscript{AC}$ reductions of at least 92.5\%. From the low-frequency IHC waveforms (bottom-most waveforms in each panel), it can be seen that the simulated amplitudes of \textsf{dau1997}, \textsf{king2019}, \textsf{relanoiborra2019}, and \textsf{osses2021} did not go below their $V\textsubscript{rest}$ (horizontal grid lines in Fig.~\ref{fig:ihc-wave}) as a result of the applied half-wave rectification process. Furthermore, \textsf{zilany2014}/\textsf{bruce2018} and \textsf{verhulst2015}  have $V\textsubscript{peak,min}$ amplitudes of $-66$ mV and $-4.7$ mV, respectively. Despite the different range in their minimum voltages, there is a strong qualitative resemblance between waveforms (green and red traces in the figure). In fact, both these models use the same type of IHC nonlinearity (compare Eqs.~17--18 from \cite{Zilany2006} with Eqs. 4--5 from \cite[][]{Verhulst2015}) and the same LP filter implementation, albeit with a different filter order and cut-off frequency (see Table~\ref{tab:detailed-info}). 

\begin{figure}
    \centering
    \includegraphics[scale=0.6]{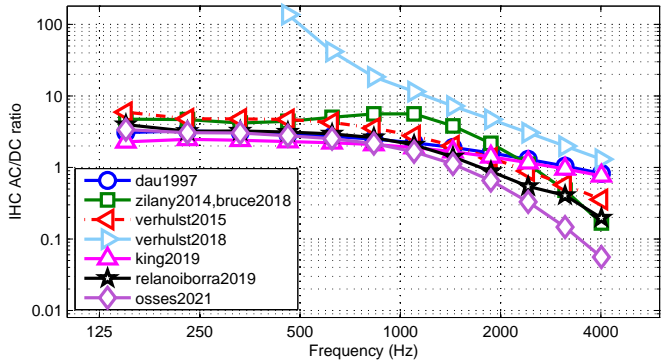}\\
    \vspace{-10pt}
    \caption{Ratio between simulated AC and DC components ($V\textsubscript{AC}/V\textsubscript{DC}$, see the text) in response to 80-dB pure tones. \textbf{Literature}: Fig.~10 from \cite{Palmer1986} and Fig.~8 from~\cite{Lopez-Poveda2006}.}
    \label{fig:ihc-ratio}
\end{figure}

The obtained AC/DC ratios are shown in Fig.~\ref{fig:ihc-ratio}, where a reduction in phase locking is given by a lower ratio. For all models, the ratio decreased with increasing frequency. All AC/DC curves, except those for \textsf{verhulst2018}, overlap well at low frequencies with ratios between 2.1 and 5.9 (below 1000 Hz), decreasing to ratios between 0.06 (\textsf{osses2021}) and 0.83 (\textsf{dau1997}) at 4013~Hz. Although the AC/DC curve for \textsf{verhulst2018} showed the highest overall ratios between 137.4 at 460 Hz down to 1.3 at 4013 Hz, due to its nearly zero DC voltages towards low frequencies (see the fairly symmetric waveforms around the horizontal grid in Fig.~\ref{fig:ihc-wave}\textbf{d}), we still observed the systematic decrease in ratio with increasing frequency. If we further focus on the AC/DC curves in the frequency range between 600 and 1000~Hz, where the phase-locking is expected to start declining  \cite{Palmer1986}, all models showed monotonically decreasing curves starting from about 833 Hz (except for \textsf{verhulst2018}, that always showed a decreasing tendency). The lowest ratios were observed for \textsf{osses2021}, followed by the similarly steep curve of \textsf{zilany2014}. Finally, a similar AC/DC curve was obtained for \textsf{relanoiborra2019} and \textsf{verhulst2015}.

\subsection{AN firing patterns}
\label{sec:comp-AN}

Simulations included AN responses to pure tones and to amplitude-modulated (AM) tones from which rate-level functions expressed as onset and steady-state responses were obtained. With these benchmarks we attempt to characterise model responses at the output of the AN synapse stage or their equivalent, with a particular interest in the phenomenon of adaptation \cite[][]{Zilany2009, Moore2013}. We comment on how adaptation is affected by the type of output of Stage~5, using either the approximations from the effective models, the average or instantaneous firing rate estimates of the phenomenological models (\textsf{zilany2014}, \textsf{bruce2018}), or the average rates of the biophysical models (\textsf{verhulst2015}, \textsf{verhulst2018}).

\subsubsection{Adaptation}

\begin{figure} 
    \centering
	\includegraphics[scale=0.51]{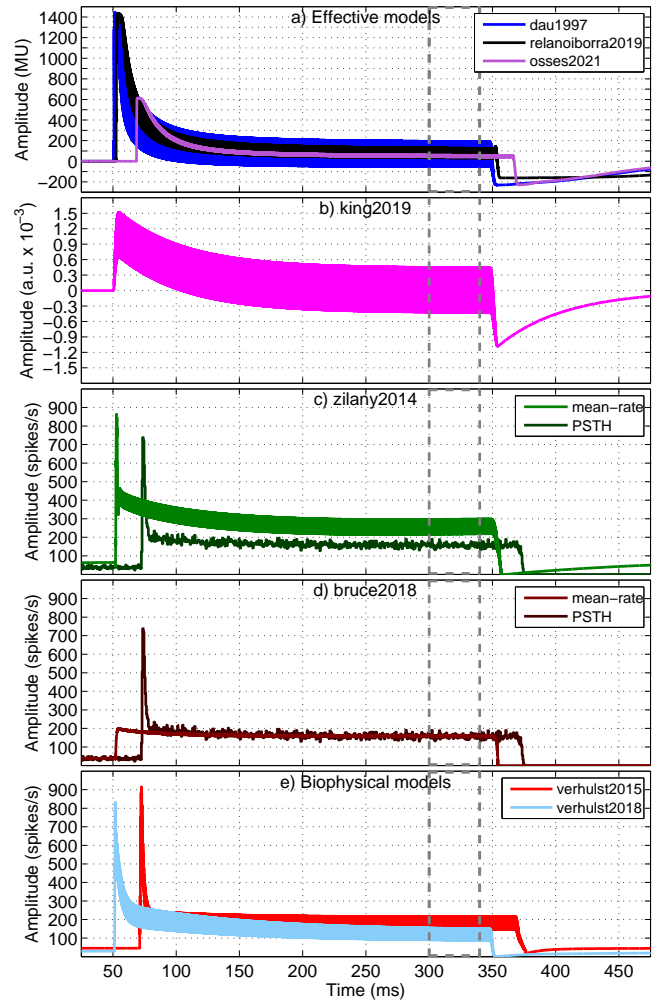}\\
    \vspace{-10pt}
    \caption{Simulated AN responses to a 4000-Hz pure tone of 70\,dB SPL. For ease of visualisation, the responses from \textsf{osses2021}, \textsf{verhulst2015}, and the PSTHs are horizontally shifted by 20 ms. \textbf{Literature}: Fig.~1 from \cite{Smith1980} and Figs.~3 and 10 from \cite{Bruce2018}.} 
    \label{fig:an-tone}
\end{figure}

To illustrate the effect of auditory adaptation, we obtained AN model responses to a 4000-Hz pure tone of 70\,dB~SPL, duration of 300\,ms, that was gated on and off with a cosine ramp of 2.5\,ms. The obtained AN responses are shown in Fig.~\ref{fig:an-tone}. All responses had an amplitude overshoot just after the tone onset which then decreased to a plateau (e.g., between 300 and 340 ms, grey dashed lines). After the tone offset ($t=350$\,ms), the AN responses showed an undershoot with decreased amplitudes that subsequently returned to their resting level. This stereotypical behaviour is related to the AN adaptation process \cite[e.g.,][]{Moore2013}. 

The waveforms from effective models using the adaptation loops (\textsf{dau1997}, \textsf{relanoiborra2019}, \textsf{osses2021}) are shown in Fig.~\ref{fig:an-tone}\textbf{a}, where their amplitudes had values between $-230.5$\,MU and 1440.2\,MU (\textsf{dau1997}), with a strong onset overshoot and a resting position at 0\,MU. For \textsf{king2019} (Fig.~\ref{fig:an-tone}\textbf{b}), a mild overshoot was observed, whose maximum amplitude (1.52$\cdot 10^{-3}$ a.u.) was higher in absolute value than that for the undershoot ($-1.08\cdot 10^{-3}$ a.u.). With an observed steady-state peak-to-peak amplitude of $0.87\cdot 10^{-3}$~a.u. \textsf{king2019} is, at this stage, the model that preserves the most temporal fine structure.

For the phenomenological models (\textsf{zilany2014} and \textsf{bruce2018}), the simulated waveforms using their two types of AN synapse outputs are shown in Fig.~\ref{fig:an-tone}\textbf{c}--\textbf{d}, based on a PSTH (dark green or brown curves) and mean-rate synapse output (light green or brown curves). The obtained PSTH and mean rate responses in \textsf{zilany2014} differ in their steady-state values (lower values for the PSTH estimate), while for \textsf{bruce2018} the difference is in their onset responses, with almost no onset adaptation in the simulated mean-rate output. For the biophysical models (Fig.~\ref{fig:an-tone}\textbf{e}), the AN synapse outputs represent mean firing rates where a stronger effect of adaptation was observed for \textsf{verhulst2018} (sky blue), with a plateau after onset that was reached after about 150 ms (at $t\approx200$\,ms) while for \textsf{verhulst2015} (red) the plateau is reached shortly after the tone onset. 

\begin{figure} 
    \centering
	\includegraphics[scale=0.48]{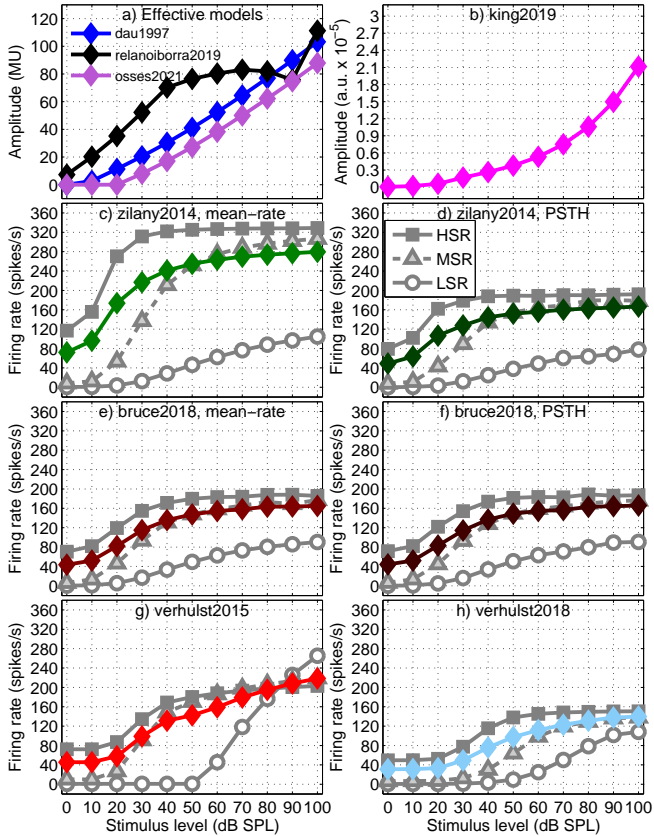}\\
    \vspace{-10pt}
    \caption{Simulated rate-level functions derived from the steady-state AN responses of 4000-Hz pure tones. For all models, average responses are shown (coloured traces). For the biophysical and phenomenological models, the responses for HSR, MSR, and LSR neurons are also shown (grey traces). \textbf{Literature}: Fig.~7 from~\cite{Bruce2018}, Fig.~5A from~\cite{Verhulst2018a}, and Fig.~3 from~\cite{Dau1996a}.} 
    \label{fig:an-discharge-rates}
\end{figure}

\begin{figure}
	\includegraphics[scale=0.48]{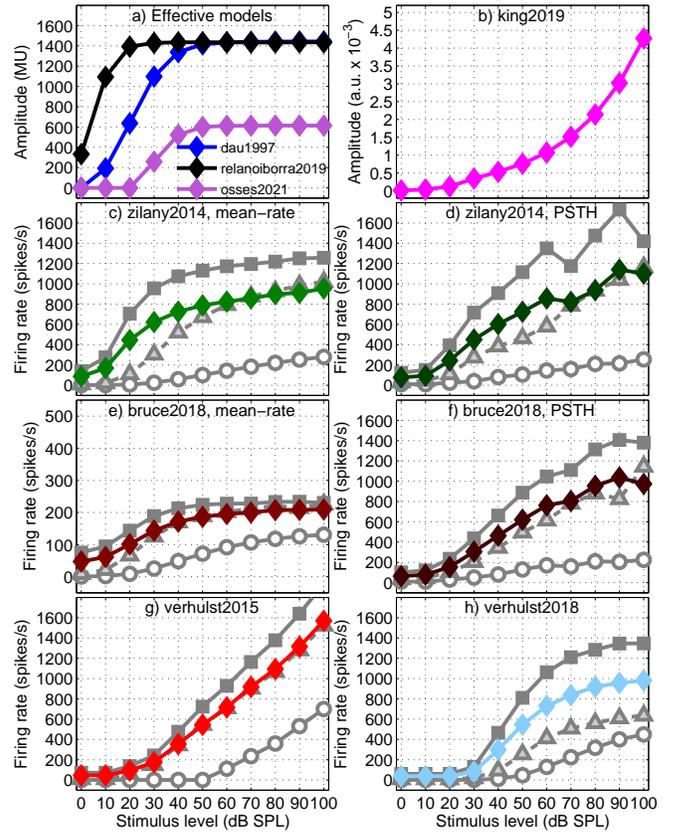}\\
    \vspace{-10pt}
    \caption{Simulated rate-level functions derived from the onset (maximum) AN responses of 4000-Hz pure tones. The colour codes and legends are as in Fig.~\ref{fig:an-discharge-rates}. \textbf{Literature}: Fig.~3 from~\cite{Smith1980}.} 
    \label{fig:an-onset-steady}
\end{figure}

\subsubsection{Rate-level functions}

Rate-level functions were simulated for a 4000-Hz pure tone presented at levels between 0 and 100\,dB~SPL with a duration of 300\,ms, gated on and off with 2.5-ms cosine ramps. The obtained results are shown in Figs.~\ref{fig:an-discharge-rates} and \ref{fig:an-onset-steady} for rate-level curves in the steady-state regime and for onset responses, respectively. For all models, average rates are shown (coloured traces) while for the phenomenological and biophysical models (panels \textbf{c--h}), the simulated response for the three types of neurons (HSR, MSR, and LSR) are shown (grey traces).

For the phenomenological and biophysical models, the discharge curves in Fig.~\ref{fig:an-discharge-rates}\textbf{c--h} tend to saturate towards higher levels, which is in line with experimental evidence \cite[e.g.,][]{Smith1980}.  
One difference between these curves is that they start to increase at slightly higher levels  for the biophysical (from $\sim$20 dB\,SPL) than for the phenomenological models (from $\sim$0 dB\,SPL). 

For the effective models (Fig.~\ref{fig:an-discharge-rates}\textbf{a--b}), with the exception of \textsf{relanoiborra2019}, the simulated rates did not show saturation as a function of level. In  \textsf{relanoiborra2019}, the simulated rates were between 70.2 and 83~MU for signal levels beyond 40\,dB. 
This saturation effect results from the combined action of the nonlinear cochlear filter (Stage~3) with the later expansion stage (Stage~5, Fig.~\ref{fig:block-diagrams}) that precedes the adaptation loops. Despite the overall lack of saturation in the evaluated effective models when looking at the steady-state outputs, a different situation is observed for the onset responses of Fig.~\ref{fig:an-onset-steady}, where the responses of the models using adaptation loops had a prominent onset saturation (\textsf{dau1997}: 1443\,MU for levels above 50\,dB; \textsf{relanoiborra2019}: 1435\,MU for levels above 30\,dB; \textsf{osses2021}: 614\,MU for levels above 50\,dB). Other interesting aspects to highlight are that: (1) almost no onset effect is observed in the mean-rate output of \textsf{bruce2018}; (2) \textsf{king2019} does not account for any type of saturation as the signal level increases (Figs.~\ref{fig:an-discharge-rates}\textbf{b}, \ref{fig:an-onset-steady}\textbf{b}). It should be noted that although  hard saturation (as in Fig.~\ref{fig:an-onset-steady}\textbf{a}) has not been experimentally observed for onset AN responses, a decrease in the rate of growth of onset-rate curves with level is expected \cite{Smith1980}, a condition that is not met in \textsf{king2019} or \textsf{verhulst2015}.


\begin{figure}
    \centering
    \includegraphics[scale=0.47]{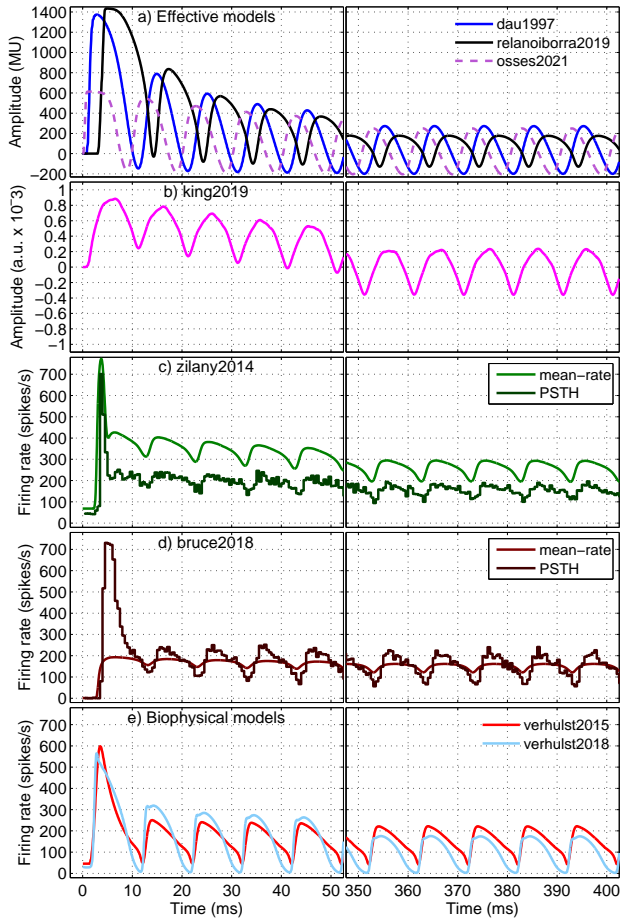}\\
    \vspace{-10pt}     
    \caption{Simulated on-frequency AN responses to a 4000-Hz AM tone of 60\,dB\,SPL (100\% modulation, $f\textsubscript{mod}=100$\,Hz). \textbf{Left}: Onset responses. \textbf{Right}: Steady-state responses. \textbf{Literature}: Fig.~12 from~\cite{Zilany2009} and Fig.~3C from~\cite{Verhulst2018a}.} 
    \label{fig:an-adaptation}
\end{figure}

\subsubsection{AM model responses}

Model responses were obtained for a 500-ms 4000-Hz pure tone that was sinusoidally modulated in amplitude (modulation index of 100\%) at a rate $f\textsubscript{mod}=100$\,Hz, presented at 60\,dB\,SPL, including up/down ramps of 2.5\,ms. The initial (0-50 ms) and later (350-400 ms) portions of the simulated responses are shown in the left and right panels of Fig.~\ref{fig:an-adaptation}, respectively. In all models, the modulation rate of 100 Hz is visible as amplitude fluctuations with the corresponding periodicity of 10 ms. In addition, adaptation was observed with stronger simulated responses immediately after the tone onset (left panels) than during the steady-state portion of the response (right panels).

For the effective models with adaptation loops (\textsf{dau1997}, \textsf{relanoiborra2019}, \textsf{osses2021}), the maximum amplitudes (Fig.~\ref{fig:an-adaptation}\textbf{a}, left) were much lower in \textsf{osses2021} than for \textsf{dau1997} and \textsf{relanoiborra2019}, due to the stronger overshoot limitation. For these models, it was also observed that their phases are not perfectly aligned due to the outer- and middle-ear filters that introduced a delay into \textsf{relanoiborra2019} (black traces run ``ahead'' the blue traces of \textsf{dau1997}), while the group-delay compensation in \textsf{osses2021} (Sec.~\ref{sec:me}) seemed to overcompensate the alignment of the simulated waveforms (purple traces run ``behind'' the blue traces). In the right panel, the dynamic range of \textsf{relanoiborra2019} (black traces) is lower than for \textsf{osses2021} and \textsf{dau1997}, which have very similar steady-state amplitudes. The reduced dynamic range in \textsf{relanoiborra2019} is mainly due to the nonlinear cochlear compression of the filter bank that interacts further with the expansion stage. In \textsf{king2019} (Fig.~\ref{fig:an-adaptation}\textbf{b}), a small effect of adaptation was observed with a maximum onset response of 0.88$\cdot 10^{-3}$ a.u. (left panel) that decreases to a local maximum amplitude of 0.24$\cdot 10^{-3}$ a.u. during the steady-state response (right panel).

The AN responses produced by \textsf{verhulst2015} and \textsf{verhulst2018} (Fig.~\ref{fig:an-adaptation}\textbf{e}) showed an overshoot reaching firing rates of 598.5 and 565.2 spikes/s, respectively. After the onset, the overshoot effect quickly disappeared in \textsf{verhulst2015}, reaching a maximum local rate of 251 spikes/s during the second modulation cycle and 222 spikes/s between 370 and 400 ms. In contrast, \textsf{verhulst2018} adapted more slowly after the onset with a maximum rate of 319 spikes/s in response to the second modulation cycle, while the response continued adapting reaching a maximum rate of 176 spikes/s between times 370 and 400~ms. 

For \textsf{zilany2014} (Fig.~\ref{fig:an-adaptation}\textbf{c}) and \textsf{bruce2018} (Fig.~\ref{fig:an-adaptation}\textbf{d}), the mean-rate and PSTH outputs are shown as lighter and darker traces, respectively. It can be observed that in \textsf{zilany2014}, the AM modulations showed a similar mean-rate and PSTH excursions of about 100\,spikes/s (Fig.~\ref{fig:an-adaptation}\textbf{b}, right: mean rates between 194 and 295\,spikes/s; PSTHs with rates between 94 and 196\,spikes/s), but the PSTHs had overall lower rates. In \textsf{bruce2018}, a greater AM fluctuation is observed for the PSTHs outputs (darker brown traces) with an excursion of 185\,spikes/s (Fig.~\ref{fig:an-adaptation}\textbf{d}, right: rates between 56 and 241\,spikes/s) compared with the 40\,spikes/s (rates between 121 and 161\,spikes/s) of its mean-rate output. Additionally, \textsf{bruce2018} showed a limited effect of adaptation in its mean-rate outputs, along with a shallower AM response in comparison to the obtained PSTH. We will not focus on the mean-rate output of this model, because (1) their authors validated the model primarily using PSTHs, recommending the use of the AN synapse output for further processing \cite{Bruce2018}, (2) the model using PSTH outputs can be used as input for subcortical processing stages from the UR EAR toolbox \cite{Carney2020a}, and (3) all the studies that we have so far identified using \textsf{bruce2018} consistently used PSTH outputs \cite[][]{Klug2020,Rahman2020}.

It should be noted that \textsf{zilany2014}, from the same model family, has been extensively validated using both mean-rate and PSTHs outputs. In fact, for studies where psychoacoustic aspects  have been investigated \cite[e.g.,][]{Maxwell2020} there is a tendency to use the mean-rate model outputs.

\subsubsection{Synchrony capture}

\begin{figure}
	\includegraphics[scale=0.48]{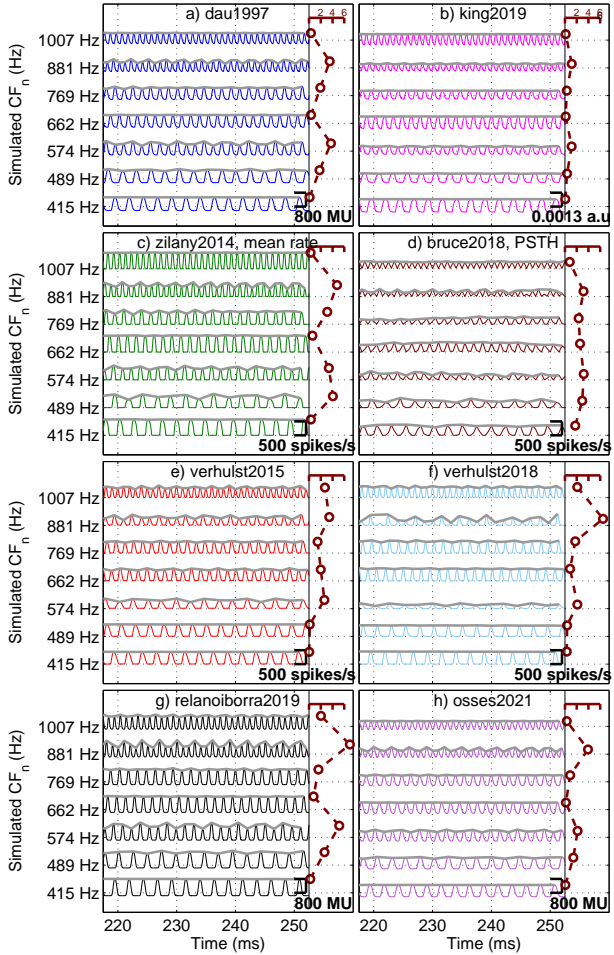}\\
    \vspace{-10pt}	
	\caption{Simulated AN responses to a complex tone with three frequency components at 414, 650, and 1000 Hz. The model simulations were obtained at on- and off-frequency CFs spaced at 1\,ERB$_N$. The thick grey lines represent the envelope of the AN responses. The maroon circle markers represent a metric that is proportional to the standard deviation of the corresponding envelope (see the text for details). \textbf{Literature}: Figs.~7--8 from~\cite{Deng1987} and Fig.~1 from~\cite{Carney2016}.} 
	\label{fig:spectrum-AN-fluctuation}
\end{figure}

Model responses were obtained for a complex tone of 50 dB SPL formed by three sinusoids of equal peak amplitude and frequencies of 414~Hz (9.6\,ERB$_N$), 650~Hz (12.6\,ERB$_N$), and 1000~Hz (15.6\,ERB$_N$). This type of complex tone with more carriers and greater range of frequencies is commonly used in studies of profile analysis \cite[e.g.,][]{Maxwell2020} and it is useful to explain an AN property named ``synchrony capture'' \cite{Bruce2003, Carney2018a} that is believed to play a relevant role in the neural coding of supra-threshold speech sounds \cite{Carney2015, Carney2018a}. When synchrony capture occurs, the neural activity in on-frequency channels is driven primarily by one frequency component in the harmonic complex, such that there are minimal fluctuations due to the fundamental-frequency envelope, while off-frequency channels exhibit fluctuating AN patterns at the fundamental frequency. To illustrate whether the evaluated models account for synchrony capture, the model outputs in response to the described complex tone were obtained for frequencies between 415 Hz ($n$=320 in Eq.\,\ref{eq:greenwood}) and 1007 Hz ($n$=245 in Eq.\,\ref{eq:greenwood}) for CFs spaced at approximately 1 ERB$_N$ ($\Delta n=12$ or 13), resulting in three on-CF and four off-CF channels. The obtained simulations are shown in Fig.~\ref{fig:spectrum-AN-fluctuation} for a 30-ms window (between 220 and 250 ms). For each waveform, a schematic metric of envelope fluctuation was obtained and shown as thick grey lines. Those envelope fluctuations were constructed by connecting consecutive local maxima that had amplitudes above the mean responses (onset excluded) of each simulated channel. Subsequently, the standard deviation of the obtained envelope estimate was (arbitrarily) divided by one thirtieth of the amplitude scales shown in the insets of each panel (e.g., divided by 800/30 MU for \textsf{dau1997}, \textsf{relanoiborra2019}, and \textsf{osses2021}). The obtained estimates were drawn as maroon circles and connected with dashed lines along the right vertical axes in Fig.~\ref{fig:spectrum-AN-fluctuation} (dimensionless scale with labels between 0 and 6, as indicated in panels \textbf{a} and \textbf{b}), where higher values indicate greater envelope fluctuation variability. The resulting envelope scale allows for a direct comparison between models. In Fig.~\ref{fig:spectrum-AN-fluctuation} it can be observed that for all models, the on-frequency channels had nearly flat envelope fluctuations. The variability estimate averaged across on-frequency bins (at 415, 662, and 1007 Hz) ranged between 0.11 (\textsf{king2019}) and 1.71 (\textsf{bruce2018}). The variability estimate across off-frequency bins (at 489, 574, 769, and 881 Hz) ranged between 0.74 (\textsf{king2019}) and 4.09 (\textsf{relanoiborra2019}, with a maximum deviation of 6.95 at CF=881 Hz in Fig.~\ref{fig:spectrum-AN-fluctuation}\textbf{g}). For all models the off-CF variability was greater than the on-CF variability, with \textsf{king2019} being the least sensitive model to code envelope fluctuations. 

\subsection{Subcortical neural processing}
\label{sec:eval_subcortical}

We show two sets of figures to schematise the subcortical processing of the evaluated models. 

\subsubsection{Modulation transfer function}

\begin{figure}
	\centering
\includegraphics[scale=0.49]{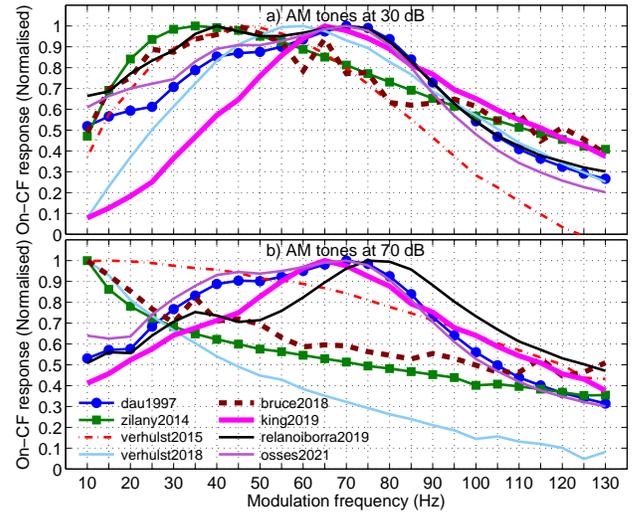}\\
    \vspace{-10pt}
    \caption{Modulation transfer functions (MTFs) of a modulation filter with a BMF$\approx80$~Hz, assessed using 1000-Hz AM tones presented at 30 (panel \textbf{a}) or 70\,dB SPL (panel \textbf{b}) that were sinusoidally modulated with $f$\textsubscript{mod} frequencies between 10 and 130 Hz. The MTFs are normalised to the maximum model response across the tested $f$\textsubscript{mod} frequencies. \textbf{Literature}: Figs. 4--6 from \cite{Krishna2000} and Figs. 1 and 4 from~\cite{Purcell2010}.}
	\label{fig:mfb-MTF}
\end{figure}

The first set of figures represents a modulation transfer function (MTF) in response to 100\% AM tones modulated at $f$\textsubscript{mod} rates between 10 and 130\,Hz (steps of 5\,Hz). The tones were centred at 1000\,Hz, had a duration of 300\,ms, included 5-ms up/down ramps, and were presented at 30 and 70\,dB SPL. For this analysis, 100\,ms in the last portion of the simulated responses were used (between times 190 and 290\,ms). The MTFs were derived from the maximum of the simulated responses. The responses were normalised to the corresponding maximum estimate over the set of tested $f\textsubscript{mod}$ values, so that the MTF of each model had a maximum value of 1. The resulting MTFs are shown in Fig.~\ref{fig:mfb-MTF}. 

The results in Fig.~\ref{fig:mfb-MTF}\textbf{a} show that the models produce bandpass-shaped MTFs with estimated BMFs between 35\,Hz (\textsf{zilany2014}) and 70\,Hz (\textsf{dau1997}, \textsf{relanoiborra2019}, and \textsf{osses2021}) that are below the theoretical BMFs (see Table~\ref{tab:detailed-info}). It is interesting to observe that the sharpest MTFs were obtained not only for \textsf{dau1997} and \textsf{osses2021} (both designed with Q=2), but also for \textsf{king2019} (which has a Q=1), while a wider tuning was observed for the remaining models, including \textsf{relanoiborra2019} (which has a Q=2). 

For the biophysical and phenomenological models, the MTFs obtained for the 70-dB AM tones (Fig.~\ref{fig:mfb-MTF}\textbf{b}) were different than those obtained for 30 dB  (Fig.~\ref{fig:mfb-MTF}\textbf{a}). For these models, the MTFs were no longer bell-shaped and seemed to act as lowpass filters, which is inline with physiological evidence indicating that regions of ``enhancement'' in MTFs of low level-signals can become regions of ``suppression'' for higher presentation levels (see, e.g., Fig.~4 from \cite{Krishna2000}). 

The effective models were more insensitive to the change in presentation levels. The only exception to this is \textsf{relanoiborra2019}, where a narrower MTF was obtained in Fig.~\ref{fig:mfb-MTF}\textbf{b} (compared with panel \textbf{a}). The models \textsf{dau1997}, \textsf{osses2021}, and \textsf{king2019} have MTFs that are qualitatively similar across presentation levels.

\subsubsection{Response to clicks of alternating polarity}

\begin{figure}
	\includegraphics[scale=0.46]{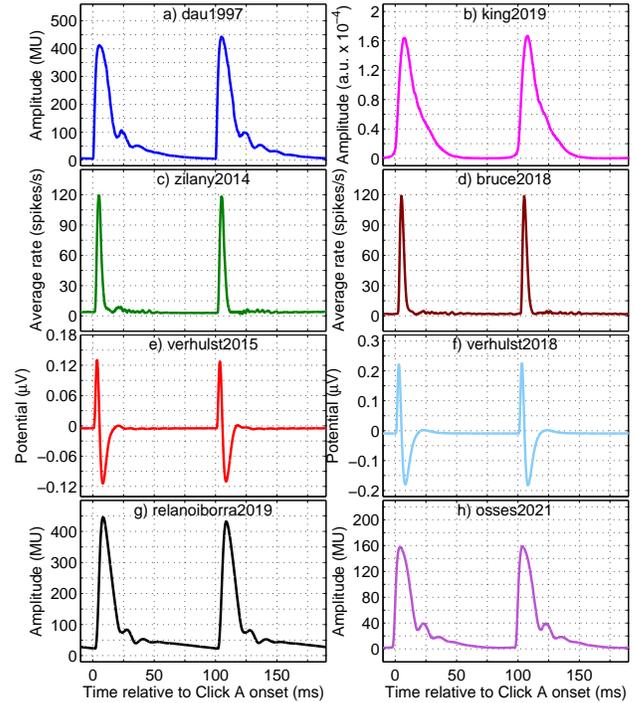}\\
    \vspace{-10pt}
	\caption{Simulated IC responses using one modulation filter (BMF$\approx$80 Hz) to a click train of alternating polarity with a total duration of 1~s, repetition rate of 10 Hz and click duration of 100 $\mu$s. Only the responses to the two last clicks are shown, whose peak-to-peak amplitudes are indicated in Table~\ref{tab:detailed-info}. \textbf{Literature}: Fig.~1 from~\cite{Schwartz1990} and Figs.~8--9 from~\cite{Picton2011-Ch08}.}
	\label{fig:mfb-click}
\end{figure}

The second set of figures focuses on simulating the response to a typical click train as used in the assessment of auditory brainstem responses (ABRs) \cite{Picton2011-Ch08}. We used a click train with a repetition rate of 10\,Hz and a duration of 1\,s (i.e., containing 10 clicks). The clicks had an alternating polarity (amplitude $A$ or $-A$) and were presented at 70 dB peak-equivalent SPL (dB\,peSPL) \cite{Laukli2015}, i.e., using $A=0.1789$\,Pa. Each individual click had a duration of 100\,$\mu$s. For this processing, the simulated outputs of Stage 6 of each model (see Fig.~\ref{fig:block-diagrams}) were averaged across CFs to obtain a broadband representation, i.e., all simulated representations were added together and then divided by the number of CFs \cite{Verhulst2015, Verhulst2018a}. This type of output can be used to derive a peak-to-peak or peak-to-trough amplitude correlate of the wave-V ABR component \cite{Picton2011-Ch08}. 

For this processing, we used the default number of CFs for the biophysical and effective models, while for \textsf{zilany2014} and \textsf{bruce2018}, 50 CFs were obtained between CF$_n=133.7$\,Hz ($n=393$, Eq.~\ref{eq:greenwood}) and CF$_n=12010$\,Hz ($n=1$), spaced at $n=8$ bins to roughly meet the number of filters from Table \ref{tab:detailed-info}. The obtained click responses are shown in Fig.~\ref{fig:mfb-click} and are illustrated for the last two clicks (of amplitudes $A$ and $-A$) of the test click train.

The biophysical models provided click responses that had positive and negative amplitudes (Fig.~\ref{fig:mfb-click}\textbf{e--f}), which was not the case for the phenomenological models that also use the SFIE model. This is because \textsf{verhulst2015} and \textsf{verhulst2018} assume that a population response can be obtained from the sum of single neuron activity (as, e.g., in \cite{ronne_modeling_2012}), with no half-wave rectification in the SFIE model (a non-explicit choice of the authors \cite{Verhulst2015, Verhulst2018a}) 
that, after scaling \cite{Verhulst2018a, Osses2019d}, results in a simplified neural representation that correlates with changes in electrical dipoles visible in scalp-recorded potentials \cite{Verhulst2018a}.

The effective models, that use the modulation-filter-bank concept, showed only positive amplitudes for all filters with BMFs $\geq10$ Hz \cite{Dau1997b} due to their envelope extraction, a phase-insensitive (``venelope'') processing \cite{Ewert2002, King2019}. For modulation frequencies below 10 Hz, the perceptual models preserve the phase information, something that is not illustrated in Fig.~\ref{fig:mfb-click} (nor in Fig.~\ref{fig:mfb-MTF}). 

Finally, the simulated peak-to-peak amplitudes in response to the last positive and negative clicks of the pulse train (ninth and tenth click, shown in Fig.~\ref{fig:mfb-click}) are shown in the entries ``Click $A$'' and ``Click $-A$'' of Table~\ref{tab:detailed-info}. From those amplitudes, it can be observed that there are models that have higher peak-to-peak amplitudes in response to positive clicks (\textsf{zilany2014}, \textsf{verhulst2015}, \textsf{bruce2018}, \textsf{relanoiborra2019}) and others where higher amplitudes are observed in  response to the clicks of negative polarity (\textsf{dau1997}, \textsf{verhulst2018}, \textsf{king2019}, \textsf{osses2021}. Although we do not discuss the significance of this polarity sensitivity, this aspect has been a matter of discussion, in particular for electrical hearing, where it has been found that evoked potentials in response to positive and negative polarity clicks represent one of the differences between humans \cite[e.g.,][]{Undurraga2010} and other mammals \cite[e.g.,][]{Ramekers2014}, whose responses are more sensitive to stimulation with clicks of negative and positive polarities, respectively.




\subsection{Computational costs}

The computational cost required to run each model was measured using the same click train as described in the previous section. Therefore, we assessed the time required to process an input signal of 1-s duration between Stages 1 to 6 of each model (Fig.~\ref{fig:block-diagrams}). This metric aims at providing a relative notion of the processing times across models. Note that some model implementations can use parallel processing, which was disabled in this evaluation. The assessment was performed on a personal computer equipped with an Intel Core i5-10210UR, 1.6-GHz processor with 16 GB of RAM memory. 

The results of the computational costs used by each model are given in the entry ``Performance'' of Table~\ref{tab:detailed-info}. The time required by the models to process one frequency channel ranged between $\sim$0.02\,s (\textsf{osses2021}, \textsf{dau1997}) and 2.5\,s (\textsf{bruce2018}). For individual frequency channels, the biophysical models (\textsf{verhulst2015} and \textsf{verhulst2018}) showed moderate calculation times between 0.3 and 0.8\,s, however, these models always require (internally) the simulation of the whole discretised cochlea with 1000 cochlear sections, independent of the number of user-requested cochlear channels (default number of 401 for the Verhulst models). This means for the current simulations, that the reported processing times of 122.9 and 319.5\,s for \textsf{verhulst2015} and \textsf{verhulst2018}, respectively, cannot be further reduced, even if the user requests the simulation of fewer CFs. In contrast, in any model based on a parallel filter bank, including~\textsf{zilany2014} and \textsf{bruce2018}, each cochlear section is independent of each other, and a user-defined number of frequency channels can be simulated, which vastly reduces the computation time for different model configurations.

Due to the long processing time of the evaluated biophysical models, their implementations include an option of parallel processing (also available in the original implementation of \textsf{bruce2018} \cite{Bruce2018}), where multiple input signals can be processed simultaneously. The number of signals that can be processed in parallel will depend on the number of threads of the host computer. As a further solution to the long processing time, Stages 2-5 of \textsf{verhulst2018} (transmission-line, IHC, and AN modules) and \textsf{bruce2018} (generating mean PSTHs) have been approximated using deep neural networks in \cite{Drakopoulos2021, Baby2021} and \cite{Nagathil2021}, respectively.

\section{Models in perspective}
\label{sec:05-perspective}

The stimuli and comparison measures used in our evaluation (Sec.~\ref{sec:04-comparison-measures}) were chosen to reflect relevant temporal and spectral properties of the models in a normal-hearing condition. Our evaluation provides an objective view, accompanied by a graphical representation of how the model responses reflect specific aspects of the hearing process in their model structure. In the following sections we provide a brief overview of the context in which each of the selected models has been used and include general recommendations for further applications.

\subsection{Applications of the evaluated auditory models}
\label{sec:discussion-features}

\textsf{Dau1997} is a monaural model that has been used to simulate a number of psychoacoustic tasks including tone-in-noise and AM detection experiments using a forced-choice paradigm \cite[e.g.,][]{Dau1996a, Dau1997b}. To enable the model for the comparison between two or more sounds, the output of Stage~6 (Fig.~\ref{fig:block-diagrams}) is used as input to a decision back-end based on a signal-detection-theory (SDT) framework, the template-matching approach. This framework, extended to adopt two templates, has been recently validated to account for the perceptual similarity between two sounds using \textsf{osses2021}~\cite{Osses2021a}.

The models \textsf{zilany2014} and \textsf{bruce2018} can account for elevated hearing thresholds due to OHC (``Cochlear gain loss'' in Stage 3) or IHC impairment (``IHC loss'' in Stage 4) \cite{Zilany2006}. The AN stage (Stage 5) includes two types of outputs: An actual spike generator and an analytical mean-rate synapse output. The spike generator has been primarily used to simulate physiological data, including the phenomenon of short- and long-term adaptation \cite{Zilany2009}. The mean-rate synapse output using \textsf{zilany2014} has been used to simulate specific psychoacoustic tasks \cite{Bianchi2019, Maxwell2020}, including speech intelligibility predictions \cite{Moncada-Torres2017}. 

The models \textsf{verhulst2015} and \textsf{verhulst2018} were initially designed to simulate otoacoustic emissions \cite{Verhulst2012} and can account for elevated hearing thresholds due to OHC impairment (``Cochlear gain loss'' in Stage 3). Furthermore, they allow to study effects of the gradual disconnection of AN fibres, known as synaptopathy, on auditory brainstem responses \cite{Verhulst2016, Verhulst2018a}. When coupled with a decision back-end, they have been used to simulate psychoacoustic performance in simultaneous tone-in-noise and high-rate AM tasks ($f\textsubscript{mod}\sim$100--120~Hz) \cite{Verhulst2018b, Osses2019b}.

The model \textsf{relanoiborra2019} can predict speech intelligibility~\cite{Relano-Iborra2019} when coupled with a decision back-end stage~\cite{Relano-Iborra2019, Joergensen2011}. Relying on the prediction power of earlier model implementations \cite{Jepsen2008, Jepsen2011}, \textsf{relanoiborra2019} should be able to (1) account for elevated thresholds based on OHC and IHC impairment \cite{Jepsen2011}, and (2) to predict a number of psychoacoustic tasks including simultaneous and forward masking and amplitude modulation \cite{Jepsen2008}. Our results showed that \textsf{relanoiborra2019} accounts well for hearing properties such as nonlinearities in the cochlear processing and auditory adaptation, including a saturation behaviour similar to that of the AN physiological models. 

The model \textsf{king2019} was designed to simulate perceptual tasks of amplitude- and frequency-modulation detection, primarily at low modulation rates ($f\textsubscript{mod}\leq20$~Hz). The model's decision back-end includes deterministic limitations (suboptimal template matching strategies) or stochastic limitations such as internal additive noise, multiplicative noise~\cite{Ewert2004}, and memory noise~\cite{Wallaert2017, Wallaert2018}. The model can be adapted to simulate hearing impairment by modifying its compression parameters (knee point and compression rate), and by increasing the bandwidth of the underlying cochlear filters. Despite the simplicity of this model---in fact one of its strengths---we have shown in this paper that the model can account for several of the comparison metrics, with the exception of the broadening of cochlear filters at higher presentation levels (Fig.~\ref{fig:noise-filterbank}), the adaptation saturation (Figs.~\ref{fig:an-discharge-rates}--\ref{fig:an-onset-steady}), and the coding of fluctuation profiles (with minimal difference in amplitude fluctuations in Fig.~\ref{fig:spectrum-AN-fluctuation}). 

\subsection{Other applications of auditory models}

Apart from the listed applications, auditory models have also been used in several other applications such as sound quality assessment \cite[e.g.,][]{Huber2006, Biberger2016, Biberger2018, Biberger2018_code}, prediction of speech intelligibility \cite[e.g.,][]{Bruce2017, Moncada-Torres2017}, and automatic speech recognition \cite[e.g.,][]{Schaedler2015}.

In the context of this special issue on binaural hearing, it is worth mentioning a number of binaural applications that rely on the evaluated monaural auditory models: The lowpass modulation filter (similar to \textsf{dau1997}) \cite[][]{Dau1996a, Dau1997b} served as the basis for a model of binaural masking that uses a decision stage based on the equalisation-cancellation theory~\cite{Breebaart2001a}. This model was later extended to predict perceptual attributes of room acoustics~\cite{Dorp2013a, Osses2017a, Osses2017b}. The model \textsf{zilany2014} has been used to predict (1)~the sensitivity to interaural time and level differences by estimating the disparity between left and right AN responses using a decision back-end based on shuffled cross-correlograms~\cite{Prokopiou2017}, and (2)~the median-plane sound localisation for various profiles of sensorineural hearing loss (OHC impairment)~\cite{baumgartner_modeling_2016}. Finally, \textsf{bruce2018} has been used to simulate the lateralisation of high-frequency stimuli in a coincidence-counting model~\cite{Klug2020}. 

\subsection{Simplified auditory representations}
\label{sec:simple}

When an auditory model is used to broaden our understanding of auditory processes \cite{meddis_computational_2010, Dau2008}, it is required that the model be as complete as possible. More details in the model often come at the price of a more computationally-expensive implementation. Such a level of detail is represented in the selected  biophysical and phenomenological models, that attempt to shed light on the mechanisms behind the cellular and neural elements included in auditory processing. On the other hand, effective models have a more epistemic status providing an intelligible but simplified representation of the process. These models can guide the design of new experiments or facilitate the development of listener-targeted products. Such model simplification, however, potentially reduces the number of effects a model can account for, leading to an actual narrowing of its application field. An example of a successful model simplification is presented in~\cite{Ewert2000}, where MTFs were simulated using only a stage of envelope extraction followed by a modulation filter bank, omitting the stages of cochlear filtering and auditory adaptation. This model, however, is not thought to predict the performance in listening conditions where the omitted model stages do play a role as it is the case (in this example) for forward-masking tasks.

Peripheral auditory models are often combined with a decision back-end module converting simulated responses into (1) a behavioural response that reflects detectability or discriminability of a sound \cite[e.g.,][]{Dau2008}, or into (2) perceptual metrics to estimate, e.g., loudness \cite{Moore1997}, perceived reverberation \cite{Dorp2013a, Osses2017a}, and sound-source localisation \cite{baumgartner_modeling_2014, McLachlan2021}. For successful simulations, the decision stage should appropriately weight the information contained in the model representations. An analysis of weighted time-frequency representations (time, audio frequency, and/or modulation frequency) can reveal what portions of the simulated responses are more relevant~\cite[e.g.,][]{Osses2021a, Joosten2016}. 

It is important to note that the simplification of auditory models based on statistical methods or machine learning processes requires a careful interpretation. While these approaches might be well suited to achieve goals such as real-time processing \cite[e.g.,][]{Drakopoulos2021} in applications of speech perception \cite[e.g.,][]{Nagathil2021} or in the prediction of evoked potentials \cite{Rahman2020}, they limit the modular comprehension of each auditory stage, especially if multiple model stages are approximated (as in \cite{Rahman2020, Nagathil2021}). 

In a recent study \cite{Rahman2020}, firing rates of cortical A1 neurons in ferrets were approximated using several time-frequency representations ranging from simple short-time Fourier transforms to more detailed models of AN synapses (including \textsf{bruce2018}) to which a linear-nonlinear (LNL) encoder was used. Based on their separately-fitted encoders, the authors concluded that cortical processing in ferrets perform a ``very simple signal transformation,'' without discussing how different the linear and nonlinear components in each of their encoders were. Despite the success of the authors in approximating neural responses in ferrets, we believe that it is difficult to know whether the ``simple transformation'' is indeed related to the underlying mechanisms of hearing (the cortical processing in ferrets) or rather is related to the complexity of operations in the fitted encoders.

\subsection{Considerations for further modelling work}

The following is a list of aspects that we recommend to keep in mind for further auditory modelling work, based on the general observations of this study:

\begin{itemize}[itemsep=.0cm,leftmargin=*]
	\item If the evaluated sounds are assumed to be reproduced via loudspeakers, or supra-aural or circumaural headphones, we recommend to use an outer-ear module as in \textsf{relanoiborra2019}, \textsf{osses2021}, or to apply an HRTF (as in \cite[][]{Bruce2003}). Although we did not evaluate this effect, such an omission implicitly assumes that the outer ear (compare the grey and black lines for \textsf{relanoiborra2019} and \textsf{osses2021} in Fig.~\ref{fig:middleear}) does not influence the coding of incoming signals in the ascending auditory pathway.
	\item The results in Figs.~\ref{fig:noise-filterbank} and \ref{fig:mfb-MTF} show that there are nonlinear interactions between model stages as a function of level and for different types of signals. This suggests that different sets of stimuli are required to characterise the behaviour of complex processes such as that of nonlinear filter banks. In other words, models may not always act as a linear time invariant (LTI) system. 
	\item In Sec.~\ref{sec:freq-sel}, we suggested a minimum number of filters for each filter bank to roughly meet a $-3$-dB filter crossing (Table~\ref{tab:detailed-info}, ``40 dB: Number of bands''). The required number of bands may vary from application to application and depend on the type of sounds that are to be simulated. This choice can be particularly critical in models where the number of bands are a free parameter (here \textsf{zilany2014} and \textsf{bruce2018}). For models that are used as front-ends to machine-learning applications, Lyon~\cite{Lyon2011} suggested a ``not-too-sparse set of channels'' with about a 50\% overlap between filters, i.e., twice the number of channels that we recommend in Table~\ref{tab:detailed-info}. It is important to keep in mind, however, that our estimation was based on model responses to white noises, which are sustained signals in time and broadband in frequency. At higher presentation levels, where nonlinear filter banks act as compressors, similar estimations using sine tones (sustained narrowband signals, as in Fig.~\ref{fig:IO-filterbank}) or clicks (transient broadband sounds) may result in a different number of required bands.
    \item Different simulation results can be expected when evaluating mean-rate and PSTH outputs of models including AN synapse stages as shown in Figs.~\ref{fig:an-tone}--\ref{fig:an-adaptation}. The particular choice of the type of output depends on the target application of the model. The spike generator is primarily used to simulate physiological data~\cite[e.g.,][]{Zilany2009, Bruce2018}, while the mean-rate synapse output is typically used to simulate specific psychoacoustic tasks \cite[e.g.,][]{Bianchi2019, Maxwell2020}.
	\item The choice of a set of stimuli to test and validate a specific model is crucial. As we stated in Sec.~\ref{sec:01-intro}, the simulation of ``unseen'' (arbitrary) sounds may produce model outputs that have not been previously validated (or at least not reported) by the model developers. Actually, an unexpected model behaviour may not be strictly related to an unseen sound, but rather to an unseen sound property. For example, the models with adaptation loops have historically had an oversensitivity to transient sounds \cite[][]{Dau1996b, Breebaart2001c, Osses2021a}, leading to model versions with limited onset responses to counteract this effect \cite{Dau1997b, Osses2021a} or have used stimuli with smoother onsets in their evaluation. 

When using large datasets, where the stimuli are split into training and validation data \cite[e.g.,][]{Harlander2014, Biberger2018, Nagathil2021}, the stimuli should contain representative samples of the relevant sound properties that the model user wishes to test. A practice like this can help to support (or not) the applicability of a specific model to sounds that may have not been even validated before, the ``unseen sounds,'' reducing (or generating awareness of) the potential limitations of the test model.
\end{itemize}

\section{Conclusions}
\label{sec:06-conclusions}

In this paper we compared eight monaural models of human auditory processing that simulate responses---with different levels of accuracy---up to the level of the inferior colliculus in the midbrain. We described and quantified the similarities and differences among model implementations and derived a minimum number of filters required for those stages to ensure the preservation of auditory information based on our estimates of frequency selectivity. 

We showed that despite the differences in model design that result in more physiologically- (biophysical and phenomenological models) or perceptually-plausible approximations (effective models), all the models can account for a number of basic hearing properties. Examples of these properties are the phase-locking reduction in inner-hair-cell processing and the phenomenon of auditory adaptation. 
Still, an in-depth understanding of each of the model stages is required when selecting a model for a specific application. We encourage future users to be explicitly aware of the specific datasets of sounds and experimental paradigms upon which their models have been evaluated, as well as of other underlying model limitations. To this end, a comparison across model implementations provides a guideline for their selection and an excellent way to challenge the capabilities of different models. 

\begin{table*}
\caption{Model configurations and numeric results. \textbf{Middle ear}: Details of frequency response of the middle-ear filters. \textbf{Cochlear filter bank}: 40~dB and 100~dB refer to filter characteristics between 160 and 8000 Hz in response to white noises of 40 and 100\,dB\,SPL, respectively. \textbf{IHC}: Parameters and frequency response of the LP filter structures. \textbf{Subcortical processing}: Theoretical and estimated BMF of the modulation filter with the closest BMF to 100 Hz and peak-to-peak  amplitude of the simulated IC outputs in response to 70-dB-peSPL clicks of positive ($A$) and negative ($-A$) (see Sec.~\ref{sec:eval_subcortical} for more details). \textbf{Performance}: Time required to process a 1-s long stimulus using each of the selected auditory models, between Stages~1 and~6 (Fig.~\ref{fig:block-diagrams}). All $f\textsubscript{cut-off}$ in this table were measured at the $-3$~dB points of the amplitude spectrum.} \vspace{-10pt}
\scalebox{0.9}{
\begin{tabular}{l|p{1.4cm}lllp{1.4cm}p{1.4cm}ll} \hline\hline
\textbf{Stage} & \multicolumn{8}{c}{Model} \\
Parameter & \textsf{dau1997} & \textsf{zilany2014} & \textsf{verhulst2015} & \textsf{verhulst2018} & \textsf{bruce2018} & \textsf{king2019} & \textsf{relanoiborra2019} & \textsf{osses2021} \\ \hline

\multicolumn{9}{l}{\textbf{Middle ear}} \\
Passband gain in Fig.~\ref{fig:middleear} (dB) & -- & $-6.0$ & $24.0$ & $18.0$ & $-6.0$ &    -- & $-66.9$ & $0.00$ \\ 
Lower $f\textsubscript{cut-off}$ (Hz)  & -- & 577.6& 601.0&601.1 &577.6&--&474.8&474.8\\
Higher $f\textsubscript{cut-off}$ (Hz) & -- &6061.9&2995.3&3993.1&6061.9&--&1230.2&1230.2\\ \hline

\multicolumn{9}{l}{\textbf{Cochlear filter bank}} \\
Reference gain in Figs.\,\ref{fig:Excitation}--\ref{fig:IO-filterbank} (dB) & $-0.6$ & $-44.1$ & $-77.9$ & $-101.9$ & $-44.1$ & $-43.9$ & $26.9$ & $-1.9$ \\ 
40\,dB: Number of filters & 34 & 51 & 59 & 52 & 51 & 34 & 36 & 34 \\ 
40\,dB: Average filter bandwidth (ERB) & 0.903 & 0.588 & 0.505 & 0.579 & 0.588 & 0.904 & 0.848 & 0.905 \\ 
100\,dB: Number of filters & 34 & 20 & 12 & 15 & 20 & 33 & 27 & 34 \\
100\,dB: Average filter bandwidth (ERB) & 0.903 & 1.57& 3.046 & 2.299 & 1.57 & 0.925 & 1.147 & 0.905 \\ \hline

\multicolumn{9}{l}{\textbf{IHC}} \\
Number of filter sections & 1 & 7 & 1 & -- & 7 & 1 & 1 & 5\\
Order of each filter section & 1 & 1 & 2 & -- & 1 & 1 & 2 & 1\\
$f\textsubscript{cut-off}$ of each filter section (Hz) & 1000 & 3000 & 1000 & -- & 3000 & 1000 & 1000 & 2000 \\

$f\textsubscript{cut-off}$ of the total filter structure (Hz) &  1000 &  966 &  642 & -- &  966 & 1000 &  1000 & 771 \\ \hline

\multicolumn{9}{l}{\textbf{Subcortical processing}} \\ 

Theoretical BMF (Hz) & 77.2 & 85.4 & 82.4 & 82.4 & 85.4 & 94.0 & 77.2 & 77.2 \\
Estimated BMF from Fig.~\ref{fig:mfb-MTF}\textbf{a} (Hz) & 70 & 35 & 45 & 60 & 45 & 65 & 70 & 70 \\

Unit of the amplitude & MU & spikes/s& $\mu$V & $\mu$V & spikes/s & a.u. & MU & MU \\
Click of amplitude A      &407.3 &  116.5 & 0.245 & 0.401  & 117.6 & 1.64$\cdot 10^{-4}$ & 423.1 &155.9\\
Click of amplitude $-$A  &437.4 &  115.8 & 0.237 & 0.407  & 117.3 & 1.67$\cdot 10^{-4}$ & 408.9 &157.3\\
\hline
\multicolumn{9}{l}{\textbf{Performance}} \\
Total time (s)          & 0.80 & 86.6 & 122.9 & 319.5   & 163.1 & 1.86 & 7.70 & 0.622 \\
Number of cochlear channels & 31  & 66  &  401  & 401   &  66 & 31 &  60 & 31 \\
Time per channel (s)      & 0.026 & 1.31 & 0.306 & 0.797 & 2.47 & 0.060 & 0.128 & 0.020 \\ \hline\hline

\end{tabular}
} 
\label{tab:detailed-info}
\end{table*}

\vspace{-10pt}
\section*{Data availability statement}

The implementations of the evaluated models (see Table~\ref{tab:sel-models}) and the model comparison (function \textsf{exp\_osses2022} to reproduce Figs.~\ref{fig:middleear}--\ref{fig:mfb-click}) are publicly available as part of the AMT toolbox (\href{https://www.amtoolbox.org}{www.amtoolbox.org}) \cite{majdak_AMT_2021} as of version 1.1.0 \cite{AMT2021}.

\vspace{-10pt}
\begin{acknowledgments}

We are grateful to several colleagues who participated in technical discussions during the writing process: Enrique Lopez-Poveda, Fotios Drakopoulos, Alessandro Altoè, Armin Kohlrausch, Thomas Biberger, and Richard Lyon. We are particularly grateful to Clara Hollomey, who provided immense support for the integration of all models into AMT~1.1. The authors AOV and LV received support from ANR (project: 17-EURE-0017), LHC received support from NIH (R01-DC010813), SV received support from the ERC project RobSpear (Grant No. 678120), and PM received support from the H2020 project SONICOM (EC Grant No. 101017743). 


\end{acknowledgments}



\bibliography{library-local}

\end{document}